\documentclass[aoas,noinfoline]{imsart}
\usepackage{lmodern}
\usepackage{amssymb,amsmath}
\usepackage{ifxetex,ifluatex}
\usepackage{fixltx2e} 
\ifnum 0\ifxetex 1\fi\ifluatex 1\fi=0 
  \usepackage[T1]{fontenc}
  \usepackage[utf8]{inputenc}
\else 
  \ifxetex
    \usepackage{mathspec}
    \usepackage{xltxtra,xunicode}
  \else
    \usepackage{fontspec}
  \fi
  \defaultfontfeatures{Mapping=tex-text,Scale=MatchLowercase}
  
\fi
\IfFileExists{upquote.sty}{\usepackage{upquote}}{}
\IfFileExists{microtype.sty}{%
\usepackage{microtype}
\UseMicrotypeSet[protrusion]{basicmath} 
}{}
\usepackage[margin=1in]{geometry}
\makeatletter
\@ifpackageloaded{hyperref}{}{%
\ifxetex
  \usepackage[setpagesize=false, 
              unicode=false, 
              xetex]{hyperref}
\else
  \usepackage[unicode=true]{hyperref}
\fi
}
\@ifpackageloaded{color}{
    \PassOptionsToPackage{usenames,dvipsnames}{color}
}{%
    \usepackage[usenames,dvipsnames]{color}
}
\makeatother
\date{}
\usepackage{natbib,upgreek,amsmath,graphicx,lineno,hyperref,subcaption}

\ifx\paragraph\undefined\else
\let\oldparagraph\paragraph
\renewcommand{\paragraph}[1]{\oldparagraph{#1}\mbox{}}
\fi
\ifx\subparagraph\undefined\else
\let\oldsubparagraph\subparagraph
\renewcommand{\subparagraph}[1]{\oldsubparagraph{#1}\mbox{}}
\fi

\begin{document}

\begin{frontmatter}

\title{Predicting Paleoclimate from Compositional Data using Multivariate Gaussian Process Inverse Prediction}
\runtitle{Inverse Prediction from Compostional Data}

\thankstext{T1}{Corresponding author}
\begin{aug}
  \author{\fnms{John R.} \snm{Tipton} \thanksref{T1,m1}\ead[label=e1]{jtipton25@gmail.com}},
  \author{\fnms{Mevin B.} \snm{Hooten} \thanksref{m2,m3,m1}},
  \author{\fnms{Connor} \snm{Nolan} \thanksref{m4}},
  \author{\fnms{Robert K.} \snm{Booth} \thanksref{m5}},
  \and
  \author{\fnms{Jason} \snm{McLachlan} \thanksref{m6}}
\affiliation{Department of Mathematical Sciences, University of Arkansas\thanksmark{m0}, Department of Statistics, Colorado State University\thanksmark{m1}, U. S. Geological Survey, Colorado Cooperative Fish and Wildlife Research Unit\thanksmark{m2}, Department of Fish, Wildlife, and Conservation Biology, Colorado State University\thanksmark{m3} Department of Geosciences, University of Arizona\thanksmark{m4}, Earth and Environmental Science Department, Lehigh University\thanksmark{m5}, and Department of Biology, University of Notre Dame\thanksmark{m6}}
\end{aug}

\begin{abstract}
Multivariate compositional count data arise in many applications including ecology, microbiology, genetics, and paleoclimate. A frequent question in the analysis of multivariate compositional count data is what underlying values of a covariate(s) give rise to the observed composition. Learning the relationship between covariates and the compositional count allows for inverse prediction of unobserved covariates given compositional count observations. Gaussian processes provide a flexible framework for modeling functional responses with respect to a covariate without assuming a functional form. Many scientific disciplines use Gaussian process approximations to improve prediction and make inference on latent processes and parameters. When prediction is desired on unobserved covariates given realizations of the response variable, this is called inverse prediction. Because inverse prediction is often mathematically and computationally challenging, predicting unobserved covariates often requires fitting models that are different from the hypothesized generative model. We present a novel computational framework that allows for efficient inverse prediction using a Gaussian process approximation to generative models. Our framework enables scientific learning about how the latent processes co-vary with respect to covariates while simultaneously providing predictions of missing covariates. The proposed framework is capable of efficiently exploring the high dimensional, multi-modal latent spaces that arise in the inverse problem. To demonstrate flexibility, we apply our method in a generalized linear model framework to predict latent climate states given multivariate count data. Based on cross-validation, our model has predictive skill competitive with current methods while simultaneously providing formal, statistical inference on the underlying community dynamics of the biological system previously not available. 
\end{abstract}


\end{frontmatter}

\section{Introduction}

A variety of data are used as proxies for climate, including tree rings,
ice cores, and pollen, with each data source requiring specialized
statistical methods. Therefore, improving statistical techniques for
reconstructing paleoclimate from proxy data are necessary to better
understand past climate. Many climate proxies are non-negative
multivariate observations that occur on the unit simplex and arise when
there is interest in explaining the proportion of a total. These types
of data are called compositional count data and examples include fungal
assays \citep{saucedo2014diversity,grantham2015fungi}, molecular
sequence data of soil microbes \citep{lauber2009pyrosequencing}, and
paleoecological data
\citep{haslett2006bayesian,booth2008testate,paciorek2009mapping,brewer2012paleoecoinformatics,salter2012fast,parnell2015bayesian}.
In general, compositional data are \(N \times d\) counts \(\mathbf{Y}\)
where, for \(i=1, \ldots, N\), \(\mathbf{y}_i\) is the \(d\)-dimensional
composition for observation \(i\) (a given location or core depth) and
\(\sum_{j=1}^d y_{ij} = M_i\) is the total count of all species for
observation \(i\). When \(M_i\) is not informative of the total
abundance of the composition, these data are called compositional count
data and are one of the most common sources of paleoclimate proxy data.
The observed composition at site \(i\) is assumed to depend on a set of
covariates \(\mathbf{x}_i\) that are only observed in the modern period
and must be predicted for the past. In this manuscript, we develop novel
statistical methods for generating predictions of unobserved climate
\(\mathbf{x}_i\) from observed compositional count data
\(\mathbf{y}_i\).

We present a new probabilistic model framework for paleoclimate
reconstruction using compositional count data and apply our model
alongside currently used Bayesian and deterministic transfer function
methods to two compositional count datasets. Both of the application
datasets are further subdivided into two subsets: 1) the modern
calibration dataset where both the species compositions and climate
variables are known and the 2) reconstruction data where only the
species compositions are observed. The goal of paleoclimate
reconstructions is to use the modern calibration dataset to learn model
parameters and then use these parameters to predict the missing climate
variables using the reconstruction data. We focus entirely on model
performance for the modern calibration data because cross-validation is
only possible during the modern calibration period. Therefore, we focus
on the modern data only, with the assumption that models that perform
well on the modern data will generalize to the reconstruction data. The
two application datasets we consider are the compositional count of
testate amoebae species in bogs relative to water table depth (a proxy
measurement of hydroclimate) and the compositional count of pollen
grains in lake sediments relative to average July temperature.

\subsection{Testate Amoeba Data}

\begin{figure}
\centering
\begin{subfigure}{0.5\textwidth}
  \centering
\centering\includegraphics[width=1.0\linewidth]{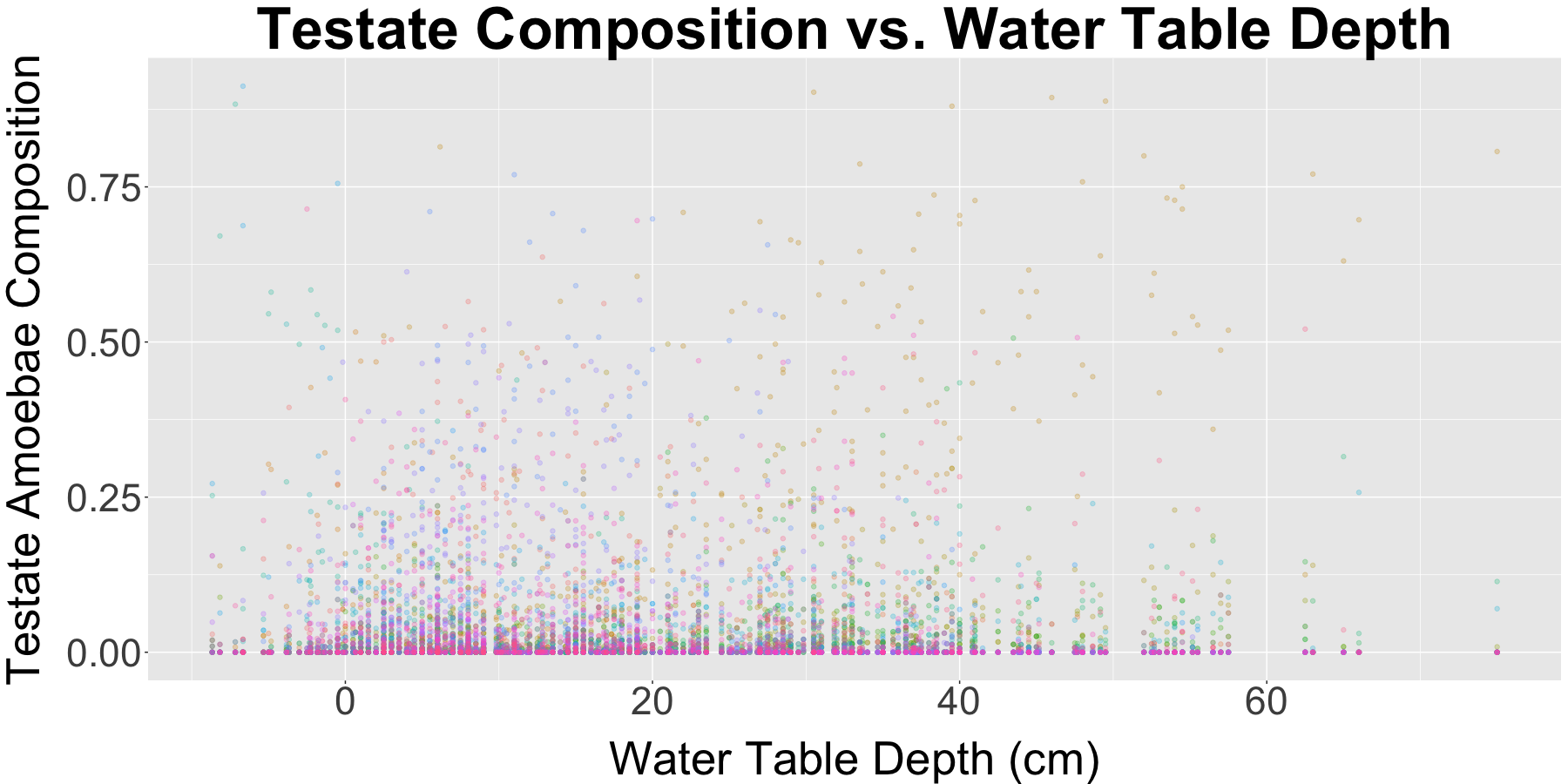}
  \caption{Testate amoeba data.}
  \label{fig:testate}
\end{subfigure}%
\begin{subfigure}{0.5\textwidth}
  \centering
  \includegraphics[width=1.0\linewidth]{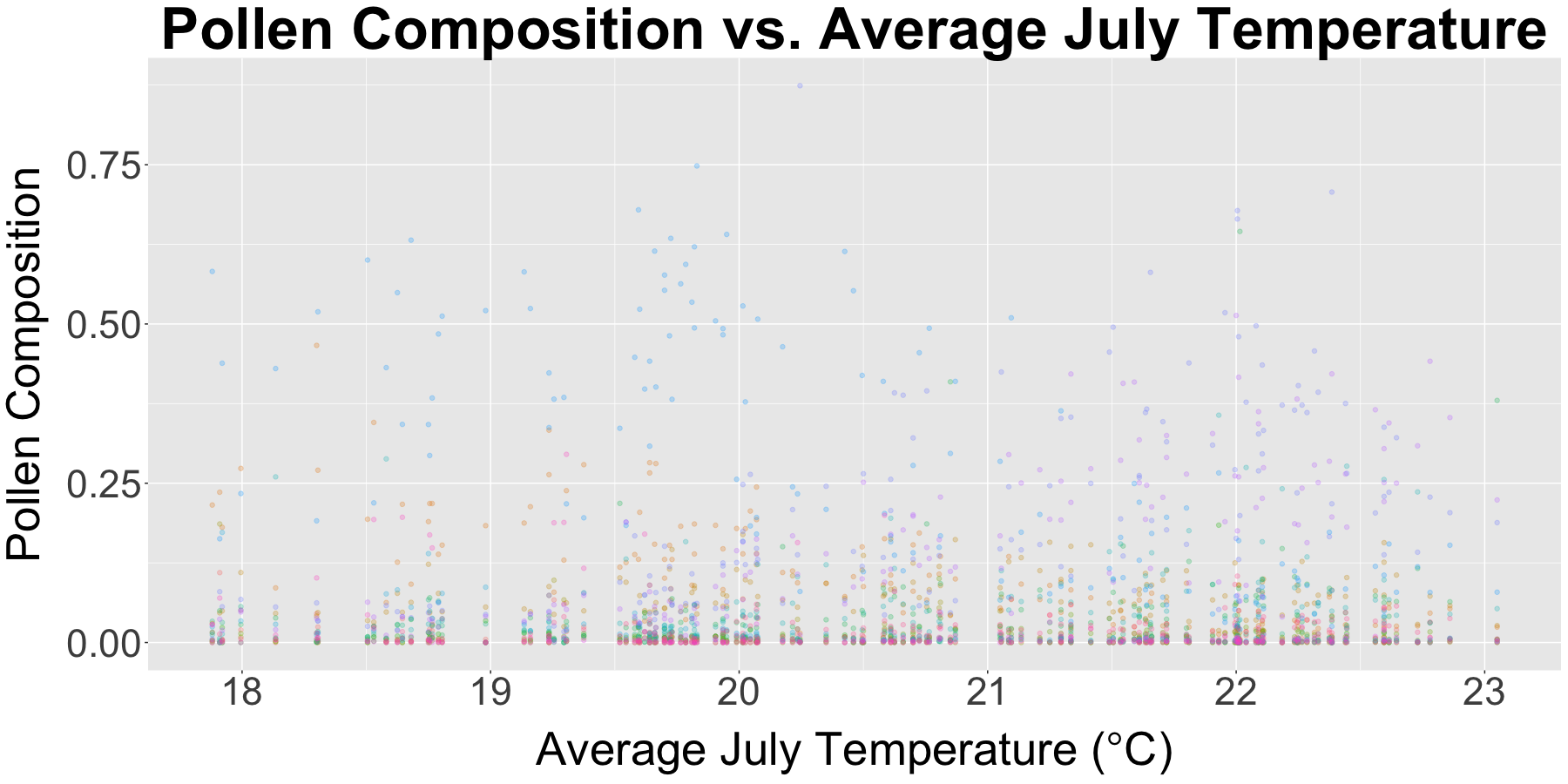}
  \caption{Pollen data.}
  \label{fig:pollen}
\end{subfigure}
\caption{Observed data with environmental covariate along the x-axis and species counts on the y-axis. Each color represents one of the $d$ observed species. Figure (a) shows the testate amoeba data and Figure (b) shows the pollen data.}
\end{figure}

Testate amoeba are a group of protists characterized by the presence of
a test (i.e., shell) which live in a range of habitats including soils,
wetlands, and peat bogs. Because testate amoebae leave behind a
decay-resistant test when they die and the tests can be used to identify
different species of amoebae, sediment cores can be used as a record of
the historical distribution of testate amoeba species. If the
distribution of testate amoebae species is sensitive to environmental
conditions, examination of the distribution of species through time can
be used to infer past climate.

Many studies have demonstrated a sensitivity of testate amoebae from
ombrotrophic peatlands to moisture conditions (i.e., water table depth)
\citep{charman2007summer,booth2010preparation,amesbury2012relationship}.
To reconstruct water table depth (in cm), we used a modern-era data set
with 356 replicates of testate amoeba assemblages including 24 species
along a water depth gradient \citep{booth2008testate}. We assumed that
the species compositions of testate amoebae at a given water table depth
are more likely to be correlated with compositions at other bog
locations with similar water table depths than nearby locations in the
same bog with different water table depths. We also made the assumption
that the response of testate amoebae species to water depth is fixed
through time. Under these assumptions, we used the data shown in Figure
\ref{fig:testate} to examine the relationship between the distribution
of testate amoebae and water table depth. While many studies have used
testate amoebae to reconstruct water depth, much remains to be learned
about the ecology of the testate amoeba communities. Important
ecological questions include how the testate amoebae species respond to
the underlying environment (water table depth) and how they interact.

\subsection{Pollen Data}

The second data source we consider is a set of pollen counts collected
at 152 sites in the Upper Midwestern United States from sediments dating
to the time of European settlement. The counts of pollen were identified
to genus level and then grouped based on shared ecological
characteristics into 16 categories. Pollen data have been studied
extensively with many reconstructions of species compositions on the
landscape \citep{dawson2016quantifying, paciorek2009mapping} and
reconstructions of paleoclimate variables from pollen data
\citep{wahl2012pollen, williams2008obtaining, haslett2006bayesian}. We
are interested in predicting temperature using the Parameter elevation
Regression on Independent Slopes Model (PRISM) 30 year normal July
temperature product in \(^\circ\)C \citep{prism} as a covariate. Figure
\ref{fig:pollen} shows the relationship between species abundances and
average July temperature. We made the assumption that there is no
spatial correlation in pollen counts conditional on July temperature,
although the model could be expanded in future work to incorporate
spatially correlated random effects.

\section{Model}

We begin with a brief introduction to the currently used determinisitic
transfer function methods for paleoclimate reconstruction using
compositional data. After introducing the most commonly used methods, we
address the shortcomings of these methods to motivate the development of
new probabilistic models. Next, we describe other Bayesian methods for
paleoclimate reconstruction and describe the assumptions of these models
that lead to the development of our model framework. We conclude this
section by introducing the novel model framework and discuss how it
solves many of the issues with previous methods.

\subsection{Transfer function methods}

The most widely used methods used in reconstructing climate from
compositional data are generally known as ``transfer functions.'' The
three transfer function methods we consider are weighted averaging (WA),
the modern analog technique (MAT), and maximum likelihood response
curves (MLRC). WA, MAT, and MLRC are methods that do not specify a joint
likelihood for the data. WA generates reconstruction predictions by
first learning parameters from the calibration data then applying those
parameters to the reconstruction data. Learning model parameters from
the calibration data occurs in three steps. First, WA estimates a
nonparametric bootstrap distribution of ``optimum'' climate giving the
prediction \(\mu_j\) for species \(j\) as
\(\hat{\mu}_j = \left(\sum_{i=1}^n y_{ij} x_i \right) / \left( \sum_{i=1}^n y_{ij} \right)\)
using the modern calibration data, where \(y_{ij}\) is the observed
proportion of species \(j\) in sample \(i\) and \(x_i\) is the climate
variable for sample \(i\). Second, out-of-bootstrap-sample estimates of
climate \(\hat{x}_{i}\) for sample \(i\) are generated by taking a
weighted sum of the species ``optima''
\(\hat{x}_{i} = \frac{\sum_{j=1}^d y_{ij} \hat{\mu}_j} {\sum_{i=1}^n y_{ij}}\)
\citep{ter1989inferring,birks1990diatoms}. The WA method induces a
strong shrinkage effect because the prediction \(\hat{x}_{i}\) is a mean
of means. The third step in generating the calibration model corrects
for this shrinkage. The shrinkage to the mean is corrected by boosting
the \(\hat{x}_{i}\)s with either a linear or spline regression between
out-of-bootstrap-sample \(x_i\) and \(\hat{x}_{i}\) resulting in a final
model that is ``de-shrunk''
\citep{birks2013diatoms,schapire1990strength}. The most common
implementations of WA in software then summarize the bootstrap sample
with the bootstrap mean and variance to generate predictions under the
assumption of a normal distribution (throwing away the information in
the bootstrap sample). For samples \(\tilde{i}\) in the reconstruction
data, predictions of paleoclimate \(\hat{x}_{\tilde{i}}\) are generated
using
\(\hat{x}_{\tilde{i}} = \frac{\sum_{j=1}^d y_{\tilde{i}j} \hat{\mu}_j} {\sum_{\tilde{i}=1}^n y_{\tilde{i}j}}\)
with uncertainties estimated from the normal approximation to the
bootstrap distribution. The uncertainty for the paleoclimate prediction
can be broken down into two sources: 1) the uncertainty from the
individual sample and 2) the average prediction error over the dataset.
In practice, the average prediction error over the dataset dominates the
uncertainty (often 95\%+ of the total uncertainty in our experience).
Therefore, the freqentist confidence interval widths are almost
independent of the observed composition.

MAT is commonly known in the statistics and machine learning literature
as \(k\) nearest neighbors. For each of the \(\tilde{i}\) samples in the
reconstruction data, a pseudo-distance is calculated to each of the
\(i=1, \ldots, n\) observations in the calibration data. There are a
variety of possible metrics, but the most commonly used in the
literature is square chord distance \citep{overpeck1985quantitative}.
The prediction \(\hat{x}_{\tilde{i}}\) for samples in the reconstruction
data is a (potentially weighted) average of the covariate of interest
for the \(k\) nearest samples in the calibration data. Confidence
intervals for the MAT prediction is generated by bootstrapping the
calibration data with uncertainties estimated using
out-of-bootstrap-sample uncertainties and assuming a normal error
distribution.

MLRC fits a curve to each marginal species component of the composition
then takes a weighted average of these curves to generate a prediction.
The idea behind MLRC is that, at certain ranges of the climate space,
different species would have higher/lower abundance. In implementation,
a logistic non-linear regression is fit to the presence/absence of each
species marginally and the model predictions are made by finding the
covariate value that gives rise to the highest probability of the
observed composition. The simpler WA was developed to borrow the ideas
of a functional response of the species to climate by reducing the
computational and numerical challenges of MLRC
\citep{ter1989inferring,birks1990diatoms}.

The transfer functions described above have been successful in
reconstructing site-level point estimates of paleoclimate; however,
transfer function methods have shortcomings. First, the transfer
function methods estimate uncertainty using out-of-sample-bootstrap
prediction errors that result in nearly uniform estimates of uncertainty
for the reconstruction samples. This is not a desirable property because
the the sample sizes used to generate the compositions change and the
uncertainty estimates do not reflect the sampling intensity. In
addition, there are compositions where the unobserved covariate can be
predicted with higher precision than other samples based on domain
knowledge, whereas current transfer functions methods are not capable of
doing this. A third shortcoming is the sensitivity to compositions with
zero counts; species with zero compositions are removed from the
calibration dataset but sometimes the reconstruction datasets can have
non-negligible counts of these species. Removing these potentially
meaningful species from the analysis could have an effect on the
estimation process. Finally, both WA and MLRC model the response of each
species marginally, ignoring the co-dependence and covariance among the
composition.

In addition, the transfer function methods each have specific issues. In
WA, species that are in high abundance dominate the prediction and the
prediction is less sensitive to less abundant species. For MAT, there is
no clear consensus on what similarity metric is best and the final
reconstruction might be sensitive to this choice. Under the
``no-analog'' setting (\citet{overpeck1992mapping} and
\citet{jackson2004modern}) where the set of compositions are different
between calibration and reconstruction datasets (i.e., there are no good
``analogs'' of the composition in the reconstruction data present in the
calibration data because the distributions of species compositions are
different), methods like WA and MAT show degraded predictive performance
(Appendix S5). What makes the no-analog problem particularily nefarious
is that one only uses the modern calibration data to evaluate model
predictive performance leading to overly optimistic results when
predictions are made using the reconstruction data. MLRC has been not
widely used due to poor poor predictive performance and large confidence
interval estimates - we include MLRC in our comparison as it is a
likelihood-based method that serves as a methodological inspiration for
our model.

Others have introduced Bayesian hierarchical models for reconstruction
of paleoclimate using compositional data that overcome some of the
issues with the transfer function methods. For instance,
\citet{vasko2000bayesian} introduced a Bayesian hierarchical model
framework (BUMMER) to model compositional count data. BUMMER assumes
that the marginal response of each species to climate follows a
symmetric, unimodal Gaussian response curve. The BUMMER model uses a
Dirichlet-multinomial likelihood (\ref{eq:DM}) and assumes the latent
random effect \(\log(\alpha_{ij})\) for observation \(i\) of species
\(j\) in (\ref{eq:like}) has the form of a Gaussian kernel

\begin{align*}
\log(\alpha_{ij}) & = \exp\left( a_j - \frac{(b_j - x_i)^2}{2 c_j^2} \right),
\end{align*}

\noindent where \(a_j\) is an offset that models baseline abundance,
\(b_j\) models the mode of the symmetric, unimodal response, and
\(c_j^2\) is a parameter that models the spread of the functional
response. In practice the data are often functional types or operational
taxonomic units (OTUs) that combine many different species, each having
one or more optimum responses that might respond asymmetrically to
climate. \citet{bhattacharya2006bayesian} used a Dirichlet process
mixture of unimodal Gaussian-shaped curves to model the response of each
species to climate; however, the Dirichlet process approach is
computationally demanding and is difficult to scale to large numbers of
species. We will demonstrate that our model framework produces
predictions that have similar skill to BUMMER when the symmetric,
unimodal assumption is valid while greatly outperforming BUMMER
predictive skill when the symmetric, unimodal assumption is violated

We introduce a novel reconstruction framework that we call the
Multivariate Gaussian Process model (MVGP) in the next section. In
constructing the model, we show how the MVGP model addresses the
criticisms of the models described above. Another strength of the MVGP
model is the ability to estimate the marginal response curves for each
species as well as the correlations among the functional responses
between species. Modeling inter-species correlations allows for further
learning about the underlying ecology of the system as well as provides
data-driven guidance on which taxa to combine together based on similar
functional responses.

\subsection{Compositional Data Model}
\label{sec:model}

Recall that compositional count data are \(N \times d\) counts
\(\mathbf{Y}\) where, for \(i=1, \ldots, N\), \(\mathbf{y}_i\) is the
\(d\)-dimensional composition for observation \(i\) (a given location or
core depth) and \(\sum_{j=1}^d y_{ij} = M_i\) is the total count of all
species for observation \(i\). Because the data are counts, we specify
the data model

\begin{align*}
\mathbf{y}_i & \sim \operatorname{Multinomial}\left(M_i, \mathbf{p}_i\right),
\end{align*}

\noindent where the \(d\)-dimensional vector
\(\mathbf{p}_i \equiv \left( p_{i, 1}, \ldots, p_{i, d} \right)'\)
represents the probabilities of sampling species \(j\) at location \(i\)
under the constraint \(\sum_{j=1}^d p_{ij}=1\).

To ensure identifiability in multinomial models, one often assumes a
reference category and fixes the value of the reference in the latent
space. A consequence of including a reference category is that the
random process \(\mathbf{p}_i\) lives in a (\(d-1\))-dimensional space.
Because we are interested in allowing each species to have its own
marginal response to the covariate while accounting for how these
responses co-vary, we do not include a fixed reference category. In
addition, visual inspection of the data and preliminary model fits
suggest the compositional data are overdispersed. Therefore, we assign a
mixing distribution for the probabilities \(\mathbf{p}_i\) using the
Dirichlet distribution, which allows modeling of each of the \(d\)
functional responses to the covariates while adding a mechanism for
overdispersion.

The Dirichlet distribution is commonly used to model distributions with
sum-to-one constraints, including probabilities. We assume

\begin{align*}
\mathbf{p}_i & \sim \operatorname{Dirichlet}\left(\boldsymbol{\alpha}_i\right),
\end{align*}

\noindent where the \(\alpha_{ij}>0\). Because the
\(\boldsymbol{\alpha}_i\) do not have a sum-to-one constraint, we model
each of the \(d\) species directly using an appropriate link function.
The choice of a conjugate prior model for \(\mathbf{p}_i\) allows us to
integrate out the latent variable \(\mathbf{p}_i\), giving rise to the
Dirichlet-Multinomial distribution

\begin{align}
\label{eq:DM}
\mathbf{y}_i & \sim \operatorname{Dirichlet-Multinomial}\left( M_i, \boldsymbol{\alpha}_i \right) \\
\nonumber & = \int  \operatorname{Multinomial}\left(\mathbf{y}_i \middle| M_i, \mathbf{p}_i\right) \operatorname{Dirichlet}\left(\mathbf{p}_i \middle| \boldsymbol{\alpha}_i\right) d\,\mathbf{p}_i 
\end{align}

\noindent where, for \(i=1, \ldots, N\) independent observations of the
\(d\)-dimensional vector \(\mathbf{y}_i\) of counts, the data model is

\begin{align*}
\left[ \mathbf{y}_i \middle| \boldsymbol{\alpha}_i\right] = \frac{\Gamma\left(\sum_{j=1}^d \alpha_{ij}\right)}{ \Gamma\left(M_i + \sum_{j=1}^d \alpha_{ij}\right)} \prod_{j=1}^d \frac{\Gamma\left(y_{ij} + \alpha_{ij}\right)} {\Gamma\left(\alpha_{ij}\right)}.
\end{align*}

\noindent To enforce the positive support of \(\boldsymbol{\alpha}_i\),
we use the log link function
\(\log\left(\boldsymbol{\alpha}_i\right) = \boldsymbol{\mu} + \boldsymbol{\zeta}_i + \boldsymbol{\varepsilon}_i\),
where \(\boldsymbol{\zeta}_i\) is a vector of random effects that is
conditional on the value of the climate covariate and
\(\boldsymbol{\varepsilon}_i\) is a vector of Gaussian error that
accounts for overdispersion and is uncorrelated with
\(\boldsymbol{\zeta}_i\).

\subsection{Gaussian Process Model}

The latent functional responses for the correlated random effect
\(\boldsymbol{\zeta}\) are modeled using Gaussian processes. For
\(i=1, \ldots, N\) replicates of the \(d\)-dimensional multivariate
latent response
\(\log\left(\boldsymbol{\alpha}_i \right) \equiv \left( \log \left( \alpha_{i1} \right), \ldots, \log \left( \alpha_{id} \right) \right)'\),
we define the correlated multivariate Gaussian processes with nugget as

\begin{align}
\label{eq:like}
\log\left(\boldsymbol{\alpha}_i \right) & = \boldsymbol{\mu} + \boldsymbol{\zeta}_i + \boldsymbol{\varepsilon}_i,
\end{align}

\noindent where \(\boldsymbol{\mu}\) is a vector of means and/or fixed
effects. For each observation \(i = 1 \ldots, N\), the marginal
distribution of the random effect
\(\boldsymbol{\zeta}_i \equiv \left( \zeta_{i1}, \ldots, \zeta_{id} \right)'\)
is mean zero Gaussian with covariance matrix \(\boldsymbol{\Sigma}\)
that accounts for the relationships among the observations
(inter-species correlations). An important scientific question is how
the \(d\) functional responses co-vary across covariate space. For
example, when modeling distributions of species, similar functional
responses to a covariate may indicate existence of a biologically
relevant functional group. Thus, inference on \(\boldsymbol{\Sigma}\)
allows for formal learning about the underlying ecological
relationships. We assume
\(\boldsymbol{\varepsilon}_i \sim \operatorname{N}\left(\mathbf{0}, \boldsymbol{\Sigma}_{\varepsilon} \right)\)
is an independent and identically distributed Gaussian error vector that
accounts for overdispersion in the data not explained by the Dirichlet
mixture of multinomial distributions and can be added or removed from
the model as necessary because it is not required for the generalized
linear model.

For each dimension \(j=1,\ldots,d\), the marginal random effect
\(\boldsymbol{\zeta}_{1:N, j} \equiv \left( \zeta_{1, j}, \ldots, \zeta_{N, j} \right)'\)
is a Gaussian process with

\begin{align}
\label{eq:marginal}
\boldsymbol{\zeta}_{1:N, j} & \sim \operatorname{N}\left(\mathbf{0}, \mathbf{C} \left( \mathbf{X}, \boldsymbol{\theta} \right) \right),
\end{align}

\noindent where
\(\mathbf{C} \left( \mathbf{X}, \boldsymbol{\theta} \right)\) is a
correlation matrix with \(i,i'\)th entry
\(\mathbf{C}_{i,i'} = c \left( \mathbf{x}_i, \mathbf{x}_{i'}, \boldsymbol{\theta} \right)\)
representing the latent correlation between observations at covariate
values \(\mathbf{x}_i\) and \(\mathbf{x}_{i'}\), where \(\mathbf{x}_i\)
is the \(i\)th row of the \(N \times q\) matrix of covariates
\(\mathbf{X}\). Hence, we assume the joint distribution of the random
effects is
\(\boldsymbol{\zeta} \equiv \left( \boldsymbol{\zeta_1}', \ldots, \boldsymbol{\zeta}_N' \right)' \sim \operatorname{N} \left( \mathbf{0}, \mathbf{C} \left( \mathbf{X}, \boldsymbol{\theta} \right) \otimes \boldsymbol{\Sigma} \right)\).
Each of the \(d\) marginal Gaussian processes have unit variance for
identifiability in the separable Kronecker product. The model can be
generalized by allowing the covariance structure to be non-separable,
but such extensions are not explored in this manuscript.

A common class of correlation functions is the Matérn class, of which
the Gaussian and exponential correlation functions are examples
\citep{stein2012interpolation}. The Matérn correlation function is given
by

\begin{align*}
c\left( \mathbf{x}_i, \mathbf{x}_{i'}, \boldsymbol{\theta} \right) & = \frac{1}{\Gamma \left( \nu \right) s^{\nu-1}} \left( \frac{2 \delta_{ii'} \sqrt{\nu}}{\rho} \right)^{\nu} \mathcal{K}_{\nu}\left( \frac{2 \delta_{ii'} \sqrt{ \nu } }{\rho} \right),
\end{align*}

\noindent where \(\delta_{ii'}\) is the Euclidean distance between
\(\mathbf{x}_i\) and \(\mathbf{x}_{i'}\),
\(\boldsymbol{\theta} = \left( \rho, \nu \right)'\), \(\rho\) is a
spatial range parameter, and \(\mathcal{K}_{\nu} \left( \cdot \right)\)
is the modified Bessel function of the second kind with smoothness
(differentiability) determined by the parameter \(\nu > 0\). By letting
\(\nu \rightarrow \infty\), the Matérn correlation function converges to
the Gaussian correlation function, and setting \(\nu = 0.5\) results in
the exponential correlation function. For the examples in this
manuscript, we use the exponential correlation function

\begin{align*}
c\left( \mathbf{x}_i, \mathbf{x}_{i'}, \theta \right) & = \exp{ \left\{ - \frac{\delta_{ii'}} {\rho} \right\}},
\end{align*}

\noindent where \(\theta = \rho\), although the framework accommodates
arbitrary correlation functions in general.

\begin{figure}
\centering
\begin{subfigure}{0.5\textwidth}
  \centering
  \includegraphics[width=1.0\linewidth]{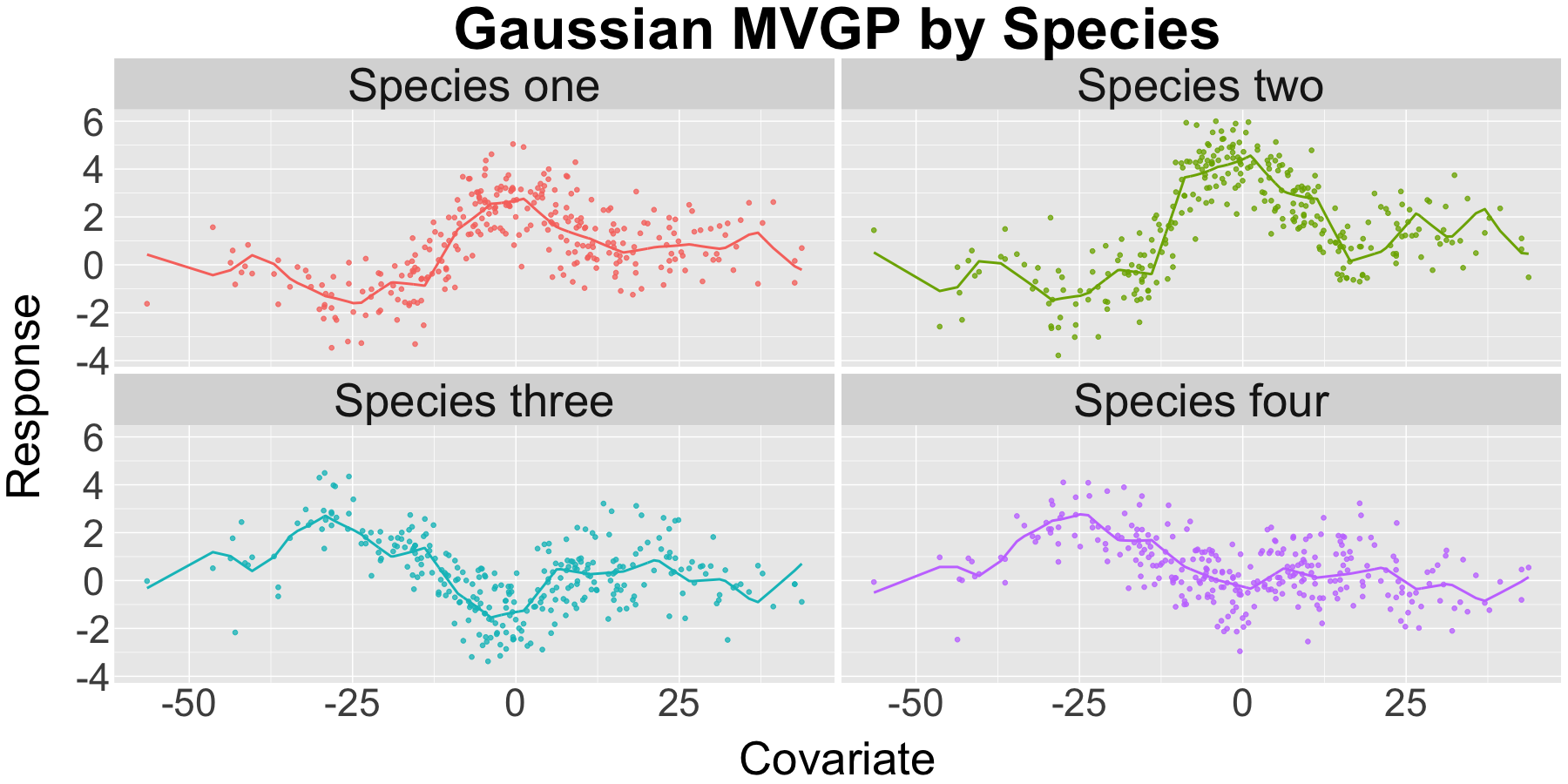}
  \caption{Simulated Gaussian process response.}
  \label{fig:sim-mvgp}
\end{subfigure}%
\begin{subfigure}{0.5\textwidth}
  \centering
  \includegraphics[width=1.0\linewidth]{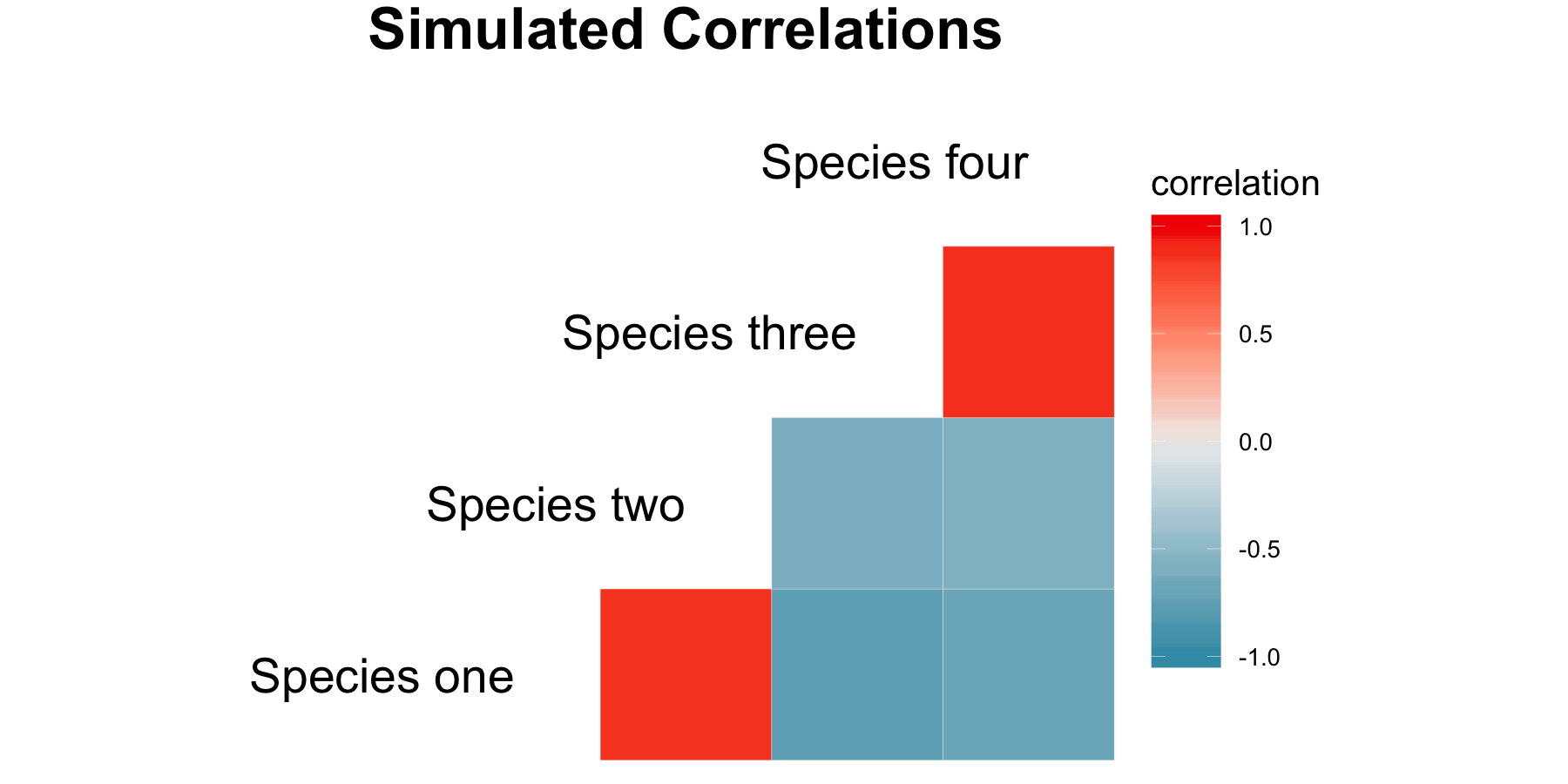}
  \caption{Simulated correlations among species.}
  \label{fig:sim-corr}
\end{subfigure}
\caption{Simulation of correlated Gaussian processes for the log-scale random effect. Notice in the left figure that species one and two show a visually similar functional response to each other with respect to the covariate value, as do species three and four. The correlations are quantified by the correlation matrix on the right showing that species one and two have a positively correlated functional response, which is seen by the similarity in functional shape of the figure on the left. Likewise, species three and four have a similar functional shape and strong positive correlation. The other pairwise comparisons have different functional shapes and associated negative correlations.}
\end{figure}

An example realization of a four-dimensional correlated Gaussian process
on the log-scale of the latent random effect (Figure \ref{fig:sim-mvgp})
shows how the log-scale random effect varies in covariate space while
allowing for the different dimensions (species) of the process to
co-vary. The flexible nature of the multivariate Gaussian process allows
for statistical learning about both the latent functional relationships
for each species and how the functional responses interact among the
species. For example, Figure \ref{fig:sim-mvgp} shows how species one
and two co-vary in their respective functional responses due to a strong
positive correlation (Figure \ref{fig:sim-corr}). We can see the
influence of the correlation on the Gaussian process realizations where
the first and second Gaussian processes (third and fourth) have very
similar shapes while the other pairwise combinations have different
shapes. The similar shapes between processes one and two (three and
four) are because the processes are highly correlated with pairwise
correlation 0.87 (0.88). The pairwise correlations of processes one and
three, one and four, two and three, and two and four are -0.76, -0.6,
-0.7, and -0.57, respectively. These negative correlations indicate that
these species have a different functional response to the underlying
covariate relative to each other. In the ecological setting, a positive
correlation in functional response could represent a relevant functional
group while a negative functional correlation could represent inhibition
or competition \citep{warton2015so,morales2015inferring}.

We rewrite \(\boldsymbol{\zeta}\) to improve computational efficiency in
our MCMC algorithm. Using the upper triangular Cholesky decomposition
\(\boldsymbol{\Sigma} = \mathbf{R}' \mathbf{R}\), define
\(\boldsymbol{\zeta} \equiv \left( \boldsymbol{\eta}_1' \mathbf{R}, \ldots, \boldsymbol{\eta}_N' \mathbf{R} \right)'\).
For \(j=1, \ldots, d\), we define

\begin{align*}
\boldsymbol{\eta}_{1:N, j} | c\left(\cdot, \cdot \middle| \boldsymbol{\theta} \right) & \sim \operatorname{N}\left(\mathbf{0}, \mathbf{C} \left( \mathbf{X}, \boldsymbol{\theta} \right) \right)
\end{align*}

\noindent as independent Gaussian processes and rewrite the latent
process (\ref{eq:like}) on the log-scale as

\begin{align}
\label{eq:likechol}
\log \left( \boldsymbol{\alpha}_i \right) & = \boldsymbol{\mu} + \mathbf{R} '\boldsymbol{\eta}_i + \boldsymbol{\varepsilon}_i,
\end{align}

\noindent where the marginal distribution of
\(\boldsymbol{\eta}_i \sim \operatorname{N}\left( \mathbf{0}, \mathbf{I} \right)\)
is multivariate standard Gaussian and the inter-species covariances are
induced by the lower Cholesky matrix \(\mathbf{R}\).

\subsection{Prior Model}
\label{sec:prior}\subsubsection{Prior on the missing covariates}

To complete the model specification, we assign prior distributions to
the remaining parameters. Because we are constructing a model for
prediction of the missing, unobserved covariates given the observed
covariates, we assign priors for the set of possible covariate values.
For spatial and temporal problems, the support of \(\mathbf{X}\) can be
restricted to the domain of interest. For predictions in covariate
space, the support of \(\mathbf{X}\) should assign positive probability
over the range of possible covariate values. We assign exchangeable
priors on each of the \(\tilde{i}=1, \ldots, \tilde{N}\) unobserved
covariate values

\begin{align}
\label{eq:Xprior}
\mathbf{x}_{\tilde{i}} & \sim  \operatorname{N}\left( \boldsymbol{\mu}_X, 1.5 \boldsymbol{\Sigma}_X \right),
\end{align}

\noindent with \(\boldsymbol{\mu}_X\) and \(\boldsymbol{\Sigma}_X\)
chosen as the sample mean and covariance of the set of observed
covariates. While this prior might not be fully Bayesian (it uses the
observed covariates as a prior for the unobserved covariates), we argue
that, \textit{a priori}, it is reasonable to assume the past values of
the covariate are similar to the current values with additional
variation in the past relative to current values. Other prior
specifications are possible as long as the prior is Gaussian (e.g., a
correlated Gaussian spatio-temporal process) because the algorithm
defined in Section \ref{sec:comp} leverages a Gaussian prior
distribution to efficiently sample the missing covariates. We assumed
the Gaussian process correlation function is isotropic, hence
\(c\left(\mathbf{x}_{\tilde{i}}, \cdot, \boldsymbol{\theta} \right)\)
depends on \(\mathbf{x}_{\tilde{i}}\) only through the set of pairwise
distances to all of the other \(\mathbf{X}\) (both observed and
unobserved). Thus, given an estimate for the distance from a location,
there are possibly many potential covariate values that can have
posterior probability mass, giving rise to multiple modes in the
posterior (i.e., the correlation function is a many-to-one function from
which we aim to generate inverse predictions). The inclusion of multiple
correlated Gaussian processes helps to constrain the multimodal
estimation but does not fully resolve the multimodal problem when signal
to noise ratio in the data model is low. The problem is also exacerbated
by a non-Gaussian likelihood associated with the compositional data. An
example of the multimodal posterior distribution of the covariate is
shown in Figures \ref{fig:testate-prediction} and
\ref{fig:pollen-prediction}. The multimodality in the conditional
posterior of \(\mathbf{x}_{\tilde{i}}\) presents a challenge which
requires an MCMC sampling algorithm that is capable of exploring
multiple modes efficiently (hence the Gaussian prior (\ref{eq:Xprior}),
see Section \ref{sec:posterior} and Section \ref{sec:comp} for details).

\subsubsection{Prior on the covariance $\boldsymbol{\Sigma}$}

The canonical inverse-Wishart prior for the covariance matrix
\(\boldsymbol{\Sigma}\) is often used for convenience because the
inverse-Wishart is conjugate to a Gaussian likelihood, but the prior has
some well-known shortcomings \citep{barnard2000modeling,o2008domain}. In
our case, the likelihood is non-Gaussian and we do not have the
computational benefits of a conjugate prior. For these reasons, we
explore an alternative prior on the covariance. We use a Cholesky
decomposition \(\boldsymbol{\Sigma} = \mathbf{R}' \mathbf{R}\) to induce
the covariance that is vague and places marginal prior probability
nearly uniformly over the space of possible correlations while assuming
the marginal variances are independent of the correlations. We follow
the separation strategy introduced in \citet{barnard2000modeling} and
model the marginal covariance
\(\boldsymbol{\Sigma} \equiv \operatorname{diag}\left( \boldsymbol{\tau} \right) \boldsymbol{\Omega} \operatorname{diag}\left( \boldsymbol{\tau} \right)\)
using a diagonal matrix
\(\operatorname{diag}\left( \boldsymbol{\tau} \right)\) with standard
deviations
\(\boldsymbol{\tau} = \left( \tau_1, , \ldots, \tau_d \right)'\) on the
diagonal, and \(\boldsymbol{\Omega}\) is an arbitrary positive definite
correlation matrix. We model the upper diagonal Cholesky decomposition
\(\mathbf{R}_{\Omega}\) of the correlation matrix
\(\boldsymbol{\Omega} = \mathbf{R}_{\Omega}' \mathbf{R}_{\Omega}\)
directly, reducing computational cost by avoiding computation of the
Cholesky decomposition
\citep{pourahmadi1999joint,pourahmadi2000maximum,smith2002parsimonious}.
Thus, we place a prior directly on the upper diagonal Cholesky matrix
\(\mathbf{R} = \mathbf{R}_{\Omega} \operatorname{diag} \left( \boldsymbol{\tau} \right)\)
by modeling \(\mathbf{R}_{\Omega}\) and \(\boldsymbol{\tau}\).

Following \citet{lewandowski2009generating} and the \citet{stan2016}, we
construct a prior for the Cholesky decomposition of
\(\mathbf{R}_{\Omega}\) with upper diagonal elements constructed from
\(B={d \choose 2}\) random variables
\(\boldsymbol{\phi} \equiv \left( \phi_1, \ldots, \phi_B \right)'\)
whose marginal support is the interval \(\left(-1, 1 \right)\). We model
the \(\phi_b\) as independent beta random variables where, for
\(b=1, \ldots, B,\)
\(\phi_b \sim 2 \operatorname{Beta}\left(\psi_b, \psi_b \right) - 1\) is
a beta distribution scaled to the interval \(\left(-1, 1 \right)\).
\citet{lewandowski2009generating} call \(\phi_b\) partial correlations
in a regular vine and arrange \(\boldsymbol{\phi}\) in an upper
triangular matrix

\begin{align*}
\boldsymbol{\Phi} & \equiv \begin{pmatrix}
0 & \phi_1 & \phi_2 & \cdots & \phi_{B-d+2} \\
0 & 0 & \phi_3 & \cdots & \phi_{B-d+3} \\
\vdots & & & \ddots & \vdots \\
0 & 0 & 0 & \cdots & \phi_{B} \\
0 & 0 & 0 & \cdots & 0 \\
\end{pmatrix},
\end{align*}

\noindent where the \(i, j\)th elements of \(\boldsymbol{\Phi}\) (when
\(i < j\)) are interpreted as the correlation between the projections of
the \(i\)th and \(j\)th random variables on the plane orthogonal to all
variables with row index \(i'\) less than \(i\). From the partial
correlations \(\boldsymbol{\phi}\), we construct the Cholesky factor
\(\mathbf{R}_{\Omega}\) using the recursive relationship
\citep{stan2016}

\begin{align*}
R_{\Omega{i,j}} & = \begin{cases}
\hfil 0 & \mbox{if } i > j, \\
\hfil 1 & \mbox{if } i = j =1, \\
\hfil \prod_{i'=1}^{i-1} \sqrt{1 - \Phi_{i',j}^2 } & \mbox{if } 1 < i = j, \\
\hfil \Phi_{i,j} & \mbox{if } 1 = i < j, \mbox{ and } \\
\hfil \Phi_{i,j} \prod_{i'=1}^{i-1} \sqrt{1 - \Phi_{i',j}^2 } & \mbox{if } 1 < i < j, \\
\end{cases},
\end{align*}

where \(R_{\Omega{i,j}}\) and \(\Phi_{i,j}\) are the \(i,j\)th elements
of \(\mathbf{R}_{\Omega}\) and \(\boldsymbol{\Phi}\), respectively.

To complete the prior on the separation model for
\(\boldsymbol{\Sigma}\), we follow the recommendations in
\citet{gelman2006prior} and assign independent half-Cauchy priors on the
standard deviations \(\boldsymbol{\tau}\), where for
\(j = 1, \ldots, d\),
\(\tau_j \sim \operatorname{half-Cauchy}\left(0, s_j \right)\). To
sample from the half-Cauchy priors efficiently, we use the result from
\citet{armagan2011generalized} where, for \(j = 1, \ldots, d\),
\(\tau^2_j \sim \operatorname{Gamma}\left( \frac{1}{2}, \lambda_j \right)\)
with mixing parameter
\(\lambda_j \sim \operatorname{Gamma}\left( \frac{1}{2}, s_j^2 \right)\)
inducing a half-Cauchy\(\left(0, s_j \right)\) distribution on the
standard deviation \(\tau_j\). The prior on the uncorrelated residual
error covariance \(\boldsymbol{\Sigma}_{\varepsilon}\) is constructed
using the same separation strategy as for \(\boldsymbol{\Sigma}\).
Depending on model performance in cross-validation in the generalized
linear mixed model framework, the uncorrelated random effect
\(\boldsymbol{\varepsilon}_i\) can be included or dropped from the
model.

\subsection{Computational Considerations}
\label{sec:posterior}

When the covariates \(\mathbf{X}\) are fixed and known, parameter
estimation for model (\ref{eq:like}) requires on the order of
\(O(d N_{total}^3)\)\footnote{Although in practical implementations, the limiting computational bottleneck is the calculation of the Cholesky decomposition of $\mathbf{C}$, which is of order $O(N_{total}^3 / 3)$ and must be repeated for each of the $d$ Gaussian processes.}
operations, with the computationally costly inversion and determinant of
the \(N_{total} \times N_{total}\) dimensional matrix \(\mathbf{C}\)
dominating the computation of the likelihood, where
\(N_{total} = N + \tilde{N}\) is the total number of locations. If
\(N_{total}\) is large, inversion of this matrix alone is
computationally challenging. For the inverse problem, where the goal is
estimation of a set of \(\tilde{N}\) unknown covariates
\(\mathbf{x}_{\tilde{i}}\), the computational cost is prohibitive for
even small \(N_{total}\), because each MCMC iteration (or optimization
step) over the \(\tilde{N}\) missing covariates is \(O(d N_{total}^3)\)
and needs to be repeated for each of the \(\tilde{N}\) unobserved
covariates resulting in the prohibitive computational cost of
\(O(\tilde{N} d N_{total}^3)\). Thus, there is need for a
computationally tractable approach for estimation under a multivariate
Gaussian process model. One approach is to use a rank one Cholesky
update on the full rank correlation matrix, which updates the Cholesky
of the correlation matrix \(\mathbf{C}\) given a single row change in
\(\mathbf{X}\). The rank one Cholesky update has the much lower
computational cost of \(O(d N_{total}^2)\) to update the unobserved
covariates because this operation uses sums of outer products of
vectors, but these operations must be repeated \(\tilde{N}\) times
resulting in total computational complexity of
\(O(\tilde{N} d N_{total}^2) + O(d N_{total}^3)\) and we found the
algorithm numerically unstable and too computationally expensive
\citep{seeger2004low}.

A number of approaches to dimension reduction for Gaussian process
models have been proposed in the literature when the goal is traditional
prediction (called Kriging in the geostatistical literature), with
respective advantages and disadvantages. \citet{lindgren2011explicit}
proposed approximating the Gaussian process with a Markov random field
by transforming the data to a regular lattice. The Gaussian Markov
Random Field approach requires the space over which the Gaussian process
occurs to be known, which is not appropriate for our problem. It may be
possible to allow for changing locations on the lattice, but the
computational difficulty in implementing this method is substantial.
Another approach for reducing computational cost in Gaussian process
models is to approximate the likelihood in the spectral domain
\citep{fuentes2007approximate,paciorek2007computational,stein2012interpolation}.
Spectral methods are difficult to implement for the traditional
prediction problem, let alone for inverse prediction, because they
require expert tuning and require the space over which the random
process occurs to be known. Another class of methods can be described as
local approximations and express the joint likelihood as a product of
conditional distributions. Examples of local likelihood methods include
the nearest neighbor Gaussian process \citep{datta2016hierarchical} and
the compositional likelihood introduced by
\citet{vecchia1988estimation}. These methods suffer from issues in the
inverse prediction problem because the meaning of local is ill-defined
when the values over which the Gaussian process occurs are unknown.

A final class of methods for approximating Gaussian processes are linear
combinations of stochastic processes, including kernel convolutions,
wavelet regression, and other methods that can broadly be described as
basis function expansions
\citep{higdon2002space,kammann2003geoadditive,banerjee2008gaussian,cressie2008fixed,nychka2015multiresolution,hefley2017basis}.
Basis function methods allow the random effect to be separate from the
space over which the Gaussian process occurs by interpolating from a set
of fixed locations, commonly referred to as knots. Instead of the
location information residing in the correlation function, the
information about location occurs in the basis expansion. Thus, any
computation that depends on the unknown covariates requires matrix
multiplication instead of matrix inversion. Although there are many
different basis function models, we choose the predictive process
\citep{banerjee2008gaussian}. The predictive process approximation to
the Gaussian process has the properties that the parameters of the
low-rank process have the same interpretation as the parameters of the
full-rank process, and the predictive process is the best low-rank
approximation of order \(\ell\) in terms of minimizing Kullback-Leibler
divergence \citep{csato2002gaussian}. Another advantage of the
predictive process is that one can implement our proposed model for any
arbitrary positive definite correlation function, allowing for a broad,
flexible class of Gaussian process models where inverse prediction is
possible.

By assuming the Gaussian process can be represented by a low-rank
predictive process approximation of order \(\ell\), we approximate the
marginal model for the random effect (\ref{eq:marginal}) as

\begin{align*}
\boldsymbol{\eta}_{1:N, j} & \approx \mathbf{c}^{\star} \left( \mathbf{X}, \mathbf{X}^{\star}, \boldsymbol{\theta} \right) {\mathbf{C}^{\star} \left( \mathbf{X}^{\star}, \boldsymbol{\theta} \right)}^{-1} \boldsymbol{\eta}_{j}^{\star} \\ 
& = \mathbf{Z} \left( \mathbf{X}, \mathbf{X}^{\star}, \boldsymbol{\theta} \right) \boldsymbol{\eta}_{j}^{\star}
\end{align*}

\noindent where
\(\mathbf{C}^{\star} \left(\mathbf{X}^{\star}, \boldsymbol{\theta} \right)\)
is the correlation matrix created by evaluating the correlation function
\(c\left( \mathbf{X}^{\star}, \mathbf{X}^{\star}, \boldsymbol{\theta} \right)\)
at the \(\ell\) fixed knots \(\mathbf{X}^{\star}\),
\(\mathbf{c}^{\star} \left( \mathbf{X}, \mathbf{X}^{\star} \boldsymbol{\theta} \right) = c\left(\mathbf{X}, \mathbf{X}^{\star}, \boldsymbol{\theta} \right)\)
is the cross-correlation of the process at the covariate locations
\(\mathbf{X}\) and the knots \(\mathbf{X}^{\star}\),
\(\mathbf{Z} \left( \mathbf{X}, \mathbf{X}^{\star}, \boldsymbol{\theta} \right) = \mathbf{c}^{\star} \left( \mathbf{X}, \mathbf{X}^{\star}, \boldsymbol{\theta} \right) {\mathbf{C}^{\star} \left( \mathbf{X}^{\star}, \mathbf{X}^{\star}, \boldsymbol{\theta} \right)}^{-1}\)
is the basis expansion matrix that contains known and unknown covariate
values, and the low-rank random process defined at the knots is the
Gaussian predictive process

\begin{align*}
\boldsymbol{\eta}_{j}^{\star} & \sim \operatorname{N}\left( \mathbf{0}, \mathbf{C}^{\star} \left( \mathbf{X}^{\star}, \mathbf{X}^{\star}, \boldsymbol{\theta} \right) \right).
\end{align*}

\noindent Most importantly for the inverse problem, the location
information (known and unknown) is in the matrix
\(\mathbf{Z} \left( \mathbf{X}, \mathbf{X}^{\star}, \boldsymbol{\theta} \right)\)
rather than the correlation function. Thus, estimating the unknown
Gaussian process hyperparameters requires the
\(O\left(d \ell^3\right) << O\left(d N_{total}^3\right)\) inversion of
the reduced-rank matrix
\(\mathbf{C}^{\star} \left( \mathbf{X}^{\star}, \mathbf{X}^{\star}, \boldsymbol{\theta} \right)\)
once per MCMC iteration (this can be done using a cached Cholesky
decomposition that is reused for each of the \(\tilde{N}\) missing
covariates) and evaluation of the likelihood for each observation with
an unknown covariate is now reduced from the
\(O\left(d N_{total}^3\right)\) matrix inversion for the full model to
the \(O(d \ell^2)\) vector-matrix multiplication
\(\mathbf{z}_i \boldsymbol{\eta}^{\star}\) , where \(\mathbf{z}_i\) is
the \(i\)th row of
\(\mathbf{Z} \left( \mathbf{X}, \mathbf{X}^{\star}, \boldsymbol{\theta} \right)\).
Hence, we have reduced the computational cost from
\(O\left(\tilde{N} d N_{total}^3\right)\) to \(O\left(d \ell^3\right)\)
+ \(O\left(\tilde{N} d \ell^2 \right)\). The reduced computational cost
for estimating unknown covariates means the predictive process approach
is well suited for our model framework, despite the predictive
processes' well-known shortcomings of producing estimates that are
overly smooth at fine-scales
\citep{finley2009improving,stein2014limitations}.

Using the predictive process, we write our approximate Gaussian process
random effect as

\begin{align}
\label{eq:zeta}
\boldsymbol{\zeta}_i^{\star} & = \mathbf{R}' \left(\mathbf{z}_i \boldsymbol{\eta}^{\star} \right)'
\end{align}

\noindent and rewrite the latent random effect on the log-scale
(\ref{eq:likechol}) as

\begin{align}
\label{eq:likepp}
\log \left( \boldsymbol{\alpha}_i \right) & = \boldsymbol{\mu} + \boldsymbol{\zeta}_i^{\star} + \boldsymbol{\varepsilon}_i. 
\end{align}

\subsection{Implementation}
\label{sec:comp}

The posterior we sample from using MCMC is

\begin{align*}
\left[\boldsymbol{\eta}^{\star}, \sigma^2, \lambda_\sigma, \tau^2, \lambda_\tau, \rho, \boldsymbol{\phi}, \left\{ \mathbf{x}' \right\}_{\tilde{i}=1}^{\tilde{N}} \middle| \{ \mathbf{y}_i, i=1, \ldots, N_{total} \} \right] & \propto \prod_{i=1}^{N_{total}} \left[ \mathbf{y}_i \middle| \boldsymbol{\eta}^{\star}, \sigma^2, \tau^2, \rho \right] \left[ \boldsymbol{\eta}^{\star} \middle| \tau^2, \rho \right] \left[\sigma^2 \middle| \lambda_{\sigma^2} \right] \left[\lambda_{\sigma^2} \right] \\
& \hspace{6mm} \times \left[\tau^2 \middle| \lambda_{\tau^2} \right] \left[\lambda_{\tau^2} \right] \left[\rho \right] \left[ \boldsymbol{\phi} \right] \prod_{\tilde{i}=1}^{\tilde{N}} \left[ \mathbf{x}_{\tilde{i}} \right],
\end{align*}

\noindent which is high dimensional and multimodal. Thus, we develop a
MCMC algorithm that is fast, efficient, and can explore a multimodal
distribution effectively. The elliptical slice sampler is a highly
efficient method for sampling from parameters whose prior distribution
is multivariate Gaussian because the elliptical slice sampler has no
tuning parameters, jointly updates parameters of interest, reduces
random walk behavior, and allows for exploration of multimodal
full-conditionals \citep{murray2010elliptical}. We found the elliptical
slice sampler to be highly effective in sampling the high-dimensional
predictive process random effect \(\boldsymbol{\eta}^{\star}\) and the
multimodal unobserved missing covariates \(\mathbf{x}_{\tilde{i}}\);
elliptical slice sampling vastly outperformed importance sampling within
Gibbs and Metropolis within Gibbs algorithms
\citep{gelfand1990sampling}. For other parameters, we use an adaptive
random walk Metropolis within Gibbs algorithm, with multivariate
proposals on the log- or logit-scale if appropriate
\citep{roberts2009examples}.

All the probabilistic models were fit using 200,000 MCMC iterations,
discarding the first 50,000 iterations as burn-in and fixing the
adaptive proposal distributions for the remaining 150,000 iterations.
Each algorithm was run for four parallel chains, thinning every 150
iterations to reduce output file size resulting in 4,000 posterior
samples. Convergence was assessed using the Gelman-Rubin \(\hat{R}\)
statistic \citep{gelman1992inference}. For the predictive process, we
assign 30 knots evenly spaced over a range extending 1.5 standard
deviations of the observed climate state beyond the minimum and maximum
observed climate state. Based on repeated model fits, we found the
inference obtained from the MVGP model to be insensitive to the number
and location of knots.

To increase computational speed, we coded the algorithm in C++ using the
\texttt{RcppArmadillo} package \citep{RCppArmadillo} within the
\texttt{R} computing environment \citep{R-Core-Team2016}. We fit
competing models in \texttt{R} using the \texttt{rioja} package
\citep{rioja}. Supplementary information, including data and code, can
be found in the attached appendices and at
\url{http://github.com/jtipton25/compositional-inverse-prediction}.

\section{Empirical Model Evaluation}

We evaluate the performance of the model framework using a simulation
study as well as a cross-validation experiment using the two
representative compositional count datasets. For paleoclimate proxy
databases, there are often one or two environmental covariates that are
measured in common across different studies. Therefore, we focus on the
predictive performance of the model framework to a single covariate
variable in what follows. To explore the impact of model assumptions on
predictive performance we consider two simulation experiments. For the
first experiment, we simulated data from the BUMMER model. The second
experiment simulated data under the MVGP model. For each of these
datasets, we can compare the estimates of the parameters from the
equivalent simulated model to demonstrate that each model (BUMMER and
MVGP) is capable of recovering the underlying simulated parameters. In
addition, we can compare the influence of the symmetric, unimodal
assumption of BUMMER relative to MVGP in simulated data. The
simulation-based experiments evaluate predictive performance using
out-of-sample test data simulated from the model under consideration. We
also estimate the missing covariates using the transfer-function methods
and compare with predictions generated using the proposed model. For
comparisons, we also fit a simplified generalized additive model (GAM)
version of the MVGP model using a B-spline approximation to the latent
functional response with a Dirichlet-multinomial likelihood.

For the cross-validation experiments, we evaluate model performance
using the calibration testate amoeba and pollen data. Model performance
is assessed by comparing predictive skill using \(k\)-fold cross
validation for each of the candidate models. To balance computation time
and to avoid model stability issues by holding-out too much data, we
cross-validated over 12-fold hold-out data sets chosen at random. All of
the empirical model experiments do not include the additional
overdispersion term \(\boldsymbol{\varepsilon}_i\) as inclusion of this
term did not improve predictive skill in cross-validation.

\subsection{Model evaluation}

We evaluated the reconstructions using mean square prediction error
(MSPE), mean absolute error (MAE), empirical 95\% coverage for either
central Bayesian 95\% credible intervals or 95\% frequentist confidence
intervals, and the continuous ranked probability score (CRPS)
\citep{gneiting2011making}. One desirable property of a scoring rule is
propriety. A proper scoring rule is one which, under expectation,
selects the optimal predictive model. A strictly proper scoring rule is
one which, under expectation, selects the optimal predictive model and
no others. MSPE (MAE) has the advantage of being widely used, easy to
understand, and easy to implement, but is not a strictly proper scoring
rule for arbitrary likelihoods. To understand why MSPE (MAE) is not
strictly proper, one can envision two models that yield the same
predictive mean but different predictive variances. The best model is
the one with empirical coverage closest to the nominal rate, but MSPE
(MAE) is unable to distinguish between the two predictions and is thus
not strictly proper. However, MSPE (MAE) is proper because the best
model has the lowest score, on average. To ensure that our scores are
proper, we define the point forecast as the mean (median) of the
predictive distribution when using MSPE (MAE). Instead of looking at
Bayesian credible interval or frequentist confidence interval coverage
and MSPE or MAE jointly, one can use a strictly proper scoring rule that
integrates this idea formally. CRPS is a scoring rule that rewards
predictions that are close to the true value while also rewarding proper
estimation of uncertainty \citep{gneiting2007probabilistic}. The CRPS is
defined for the cumulative predictive distribution \(F\) and
out-of-sample realization \(y_{oos}\) as

\begin{align*}
CRPS\left(F, y_{oos} \right) & = \int_{-\infty}^{\infty} \left( F \left( y \right) - \mathbf{I} \left( y \geq y_{oos} \right) \right)^2 \,dy.
\end{align*}

When fitting a Bayesian model using sampling methods, including MCMC,
one can approximate the CRPS using

\begin{align*}
CRPS & = \frac{1}{2 K^2} \sum_{k=1}^K \sum_{\kappa=1}^K \left| y^{(k)} - y^{(\kappa)} \right| - \frac{1}{K} \sum_{k=1}^K \left| y^{(k)} - y_{oos} \right|,
\end{align*}

\noindent where \(y^{(k)}\) and \(y^{(\kappa)}\) are the \(k\)th and
\(\kappa\)th samples from the posterior predictive distribution. For the
nonprobabilistic WA, MAT, and MLRC methods that produce point forecasts,
the CRPS score is equivalent to MAE. \subsection{Results}

\begin{table}
\caption{Results for predicting unobserved covariate values using simulated data. Smaller MSPE, MAE, and CRPS values indicate better model performance. Central Bayesian credible interval and frequentist confidence interval coverage values closer to the nominal 95\% credible interval indicate better model performance.}
\begin{subtable}{0.45\textwidth}
\centering
\caption{Simulated BUMMER data.}
\label{tab:exp-one}
\begin{tabular}{rrrrr}
  \hline
 & CRPS & MSPE & MAE & 95\% CI coverage \\ 
  \hline
MVGP & 0.6502 & 1.4008 & 0.9105 & 95.0000 \\ 
  GAM & 0.6456 & 1.3683 & 0.9121 & 96.0000 \\ 
  BUMMER & 0.6397 & 1.3480 & 0.9154 & 93.5000 \\ 
  WA & 1.1143 & 1.8814 & 1.1143 & 98.0000 \\ 
  MAT & 1.1057 & 1.9993 & 1.1057 & 98.5000 \\ 
  MLRC & 1.6331 & 4.3291 & 1.6331 & 100.0000 \\ 
   \hline
\end{tabular}

\end{subtable}\hfill%
\begin{subtable}{0.45\textwidth}
\centering
\caption{Simulated MVGP data.}
\label{tab:exp-two}
\begin{tabular}{rrrrr}
  \hline
 & CRPS & MSPE & MAE & 95\% CI coverage \\ 
  \hline
MVGP & 0.5162 & 1.2637 & 0.6793 & 96.0000 \\ 
  GAM & 0.5394 & 1.3372 & 0.7113 & 96.5000 \\ 
  BUMMER & 0.7198 & 1.7937 & 1.0057 & 96.0000 \\ 
  WA & 1.0962 & 1.9467 & 1.0962 & 96.0000 \\ 
  MAT & 0.9413 & 1.6814 & 0.9413 & 95.0000 \\ 
  MLRC & 1.3245 & 3.3055 & 1.3245 & 95.0000 \\ 
   \hline
\end{tabular}

\end{subtable}
\end{table}

For the first simulated data experiment, the BUMMER model that assumes a
symmetric, unimodal functional response of each species to climate was
used to generate the data. Details about the simulation parameters can
be found in Appendix S1. Predictive scores in Table \ref{tab:exp-one}
show that, for data simulated under the BUMMER model, the BUMMER model
slightly outperforms MVGP and the simpler GAM across all metrics and all
the probabilistic methods outperform the transfer function methods. This
is unsurprising as the BUMMER model was used to simulate the data and is
the correct model; however, the GAM and MVGP models produce predictions
with essentially equal skill. The transfer function methods fail to
perform as well as the probabilistic models under the simulated BUMMER
data.

For the second simulation experiment, we simulated compositional count
data from the MVGP model and generated predictions using the candidate
models (Appendix S2). The results in Table \ref{tab:exp-two} show the
MVGP model performs best across all metrics with the simpler GAM model
performing nearly as well. The BUMMER model shows decreased predictive
skill because the symmetric, unimodal functional response assumption has
been violated. The transfer function methods fail to perform as well as
MVGP and GAM, with MAT outperforming WA and MLRC. The MVGP model
estimated the parameters simulated from the MVGP model accurately,
demonstrating that MVGP is useful for prediction and inference (Appendix
S2). By providing good predictions and accurately estimating the
simulated parameters in data similar to the observed data of interest,
the MVGP framework is shown to be useful for both prediction and
inference using the simulated data. From this experiment, we have shown
that the BUMMER model shows degraded predictive performance when the
assumption of a symmetric, unimodal functional response is violated.

\begin{figure}
\centering
\begin{subfigure}{0.5\textwidth}
  \centering
  \includegraphics[width=1.0\linewidth]{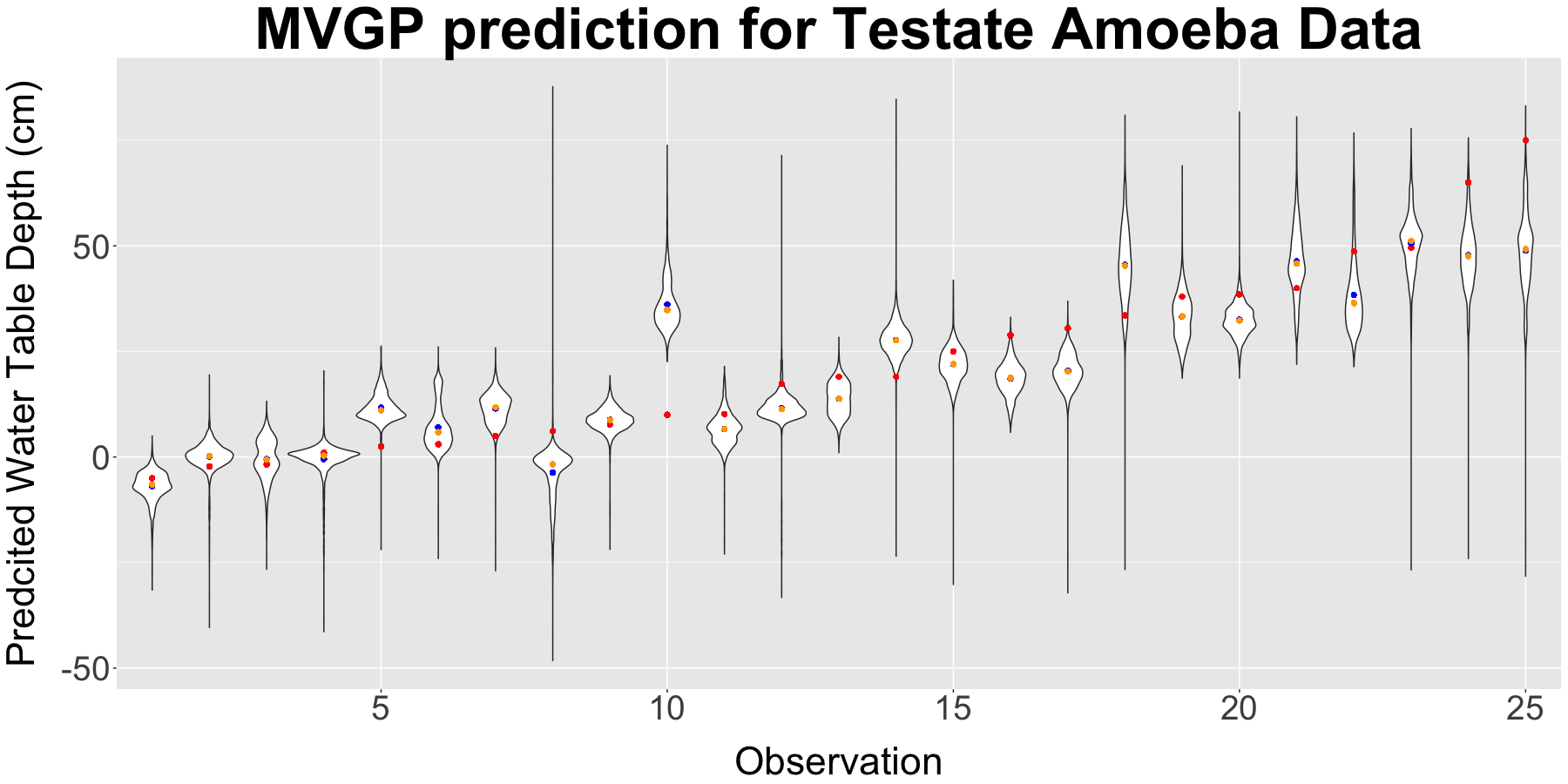}
  \caption{Held-out testate amoeba data.}
  \label{fig:testate-prediction}
\end{subfigure}%
\begin{subfigure}{0.5\textwidth}
  \centering
  \includegraphics[width=1.0\linewidth]{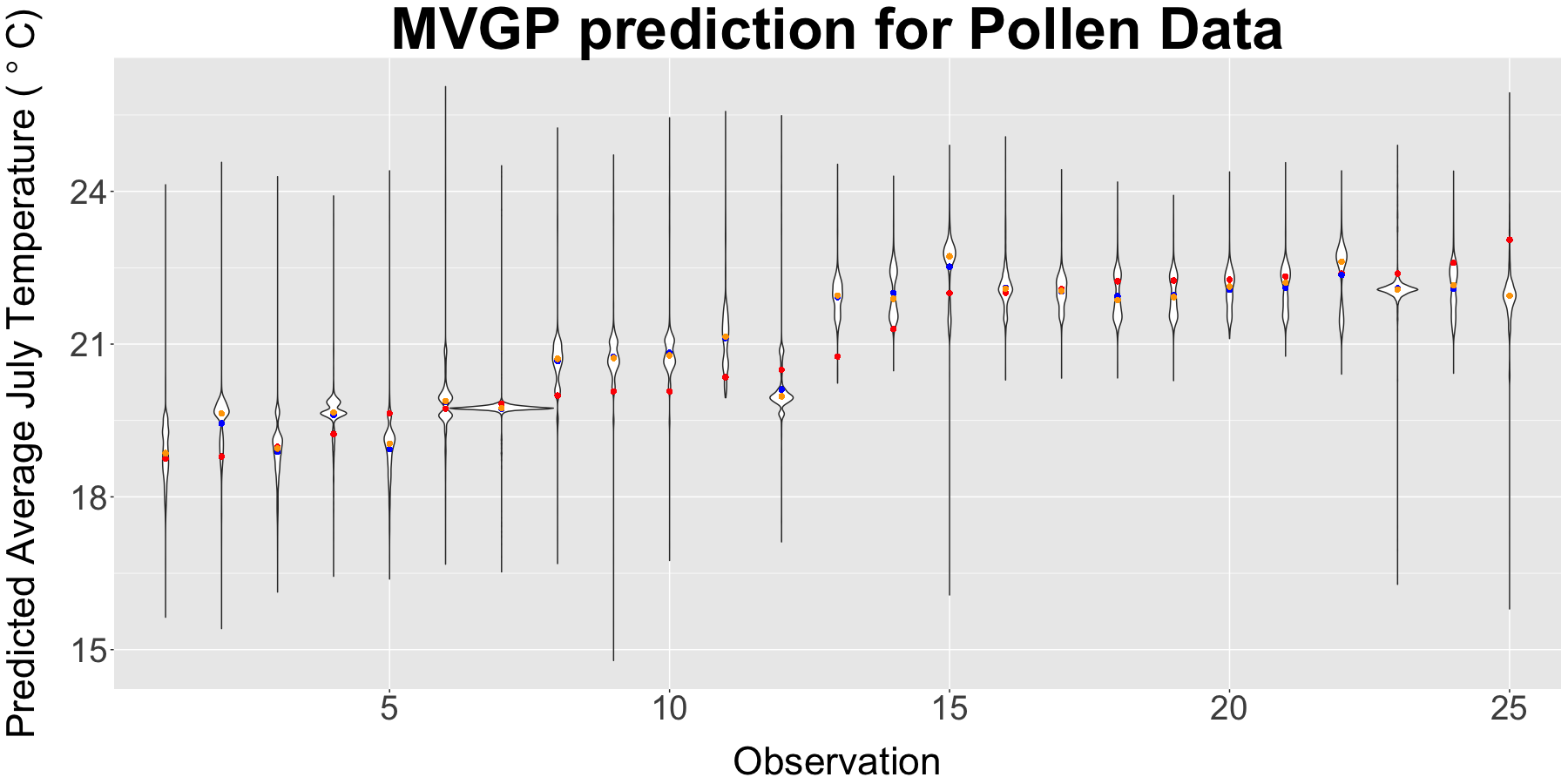}
  \caption{Held-out pollen data.}
  \label{fig:pollen-prediction}
\end{subfigure}
\caption{95\% predictive distributions for 25 held-out observations from the testate amoeba data in Figure (a) and the pollen data in Figure (b) sorted to be increasing in the held-out covariate. The dots are at the held-out true values and the violin plots represent posterior distributions. Notice that the prediction distributions show evidence of the multimodality induced by the symmetry of the correlation function and the MCMC algorithm is capable of efficiently exploring the multiple modes. In addition, the MVGP model produces posterior distributions that have varying width for different samples.}
\end{figure}

We used 12-fold cross-validation to test model performance on the
testate amoeba and pollen data to better explore the predictive
performance of the candidate models. Details for the experiments can be
found in Appendices S3 and S4 for the testate amoeba and pollen data,
respectively. The cross-validation experiment results in Table
\ref{tab:exp-testate} (testate amoeba data) and Table
\ref{tab:exp-pollen} (pollen data) demonstrate that the MVGP and GAM
models generate the best probabilistic predictions, whereas MAT produced
the best point predictions with MVGP, GAM, and WA performing similarly
on these metrics. For the CRPS metric MVGP, GAM, and BUMMER are the best
performing models in that order. The MAT produced confidence intervals
with slightly higher than expected coverage; MVGP and GAM produced
central 95\% Bayesian credible intervals with lower than expected
empirical coverage. In general, the cross-validated central Bayesian
credible interval coverage for MVGP, GAM, and BUMMER is low for the real
data sets, perhaps due to additional overdispersion that is not
accounted for in the data model. The MLRC method produced the worst
predictions across both studies, but is included because MLRC is a
functional response method like MVGP and GAM and is robust to the
no-analog problem (Appendix S5). By being competitive with currently
used predictive methods, the cross-validation experiment demonstrates
the potential of the MVGP inverse prediction framework for paleoclimate
reconstruction. The predictions shown in Figure
\ref{fig:testate-prediction} and \ref{fig:pollen-prediction} demonstrate
that the model generates reasonable predictive distributions, but the
multimodality suggests that the posterior mean (or median) implied by
use of MSPE (MAE) might not always a good description of the predictive
distribution for MVGP and GAM - hence the decreased performance on the
point prediction metrics of MSPE and MAE.

\begin{table}
\caption{Cross-validation scores for experiment three using the application data.  Smaller MSPE, MAE, and CRPS values indicate better model performance. Central Bayesian credible interval and frequentist confidence interval coverage values closer to the nominal 95\% credible interval indicate better model performance.}
\begin{subtable}{0.45\textwidth}
\centering
\caption{Testate amoeba data.}
\label{tab:exp-testate}
\begin{tabular}{rrrrr}
  \hline
 & CRPS & MSPE & MAE & 95\% CI coverage \\ 
  \hline
MVGP & 0.2959 & 0.2793 & 0.3897 & 78.3708 \\ 
  GAM & 0.2966 & 0.2933 & 0.3943 & 79.7753 \\ 
  BUMMER & 0.3073 & 0.3020 & 0.4077 & 75.8427 \\ 
  WA & 0.4014 & 0.2924 & 0.4014 & 94.3820 \\ 
  MAT & 0.3867 & 0.2340 & 0.3867 & 98.8764 \\ 
  MLRC & 0.4088 & 0.4310 & 0.4088 & 92.6966 \\ 
   \hline
\end{tabular}

\end{subtable}\hfill%
\begin{subtable}{0.45\textwidth}
\centering
\caption{Pollen data.}
\label{tab:exp-pollen}
\begin{tabular}{rrrrr}
  \hline
 & CRPS & MSPE & MAE & 95\% CI coverage \\ 
  \hline
MVGP & 0.2746 & 0.2178 & 0.3864 & 87.5000 \\ 
  GAM & 0.2828 & 0.2119 & 0.3912 & 78.9474 \\ 
  BUMMER & 0.2895 & 0.2485 & 0.4001 & 79.6053 \\ 
  WA & 0.3711 & 0.2287 & 0.3711 & 94.0789 \\ 
  MAT & 0.3486 & 0.1821 & 0.3486 & 100.0000 \\ 
  MLRC & 0.3880 & 0.2722 & 0.3880 & 96.7105 \\ 
   \hline
\end{tabular}

\end{subtable}
\end{table}

\begin{figure}
\centering
\begin{subfigure}{0.5\textwidth}
  \centering
  \includegraphics[width=1.0\linewidth]{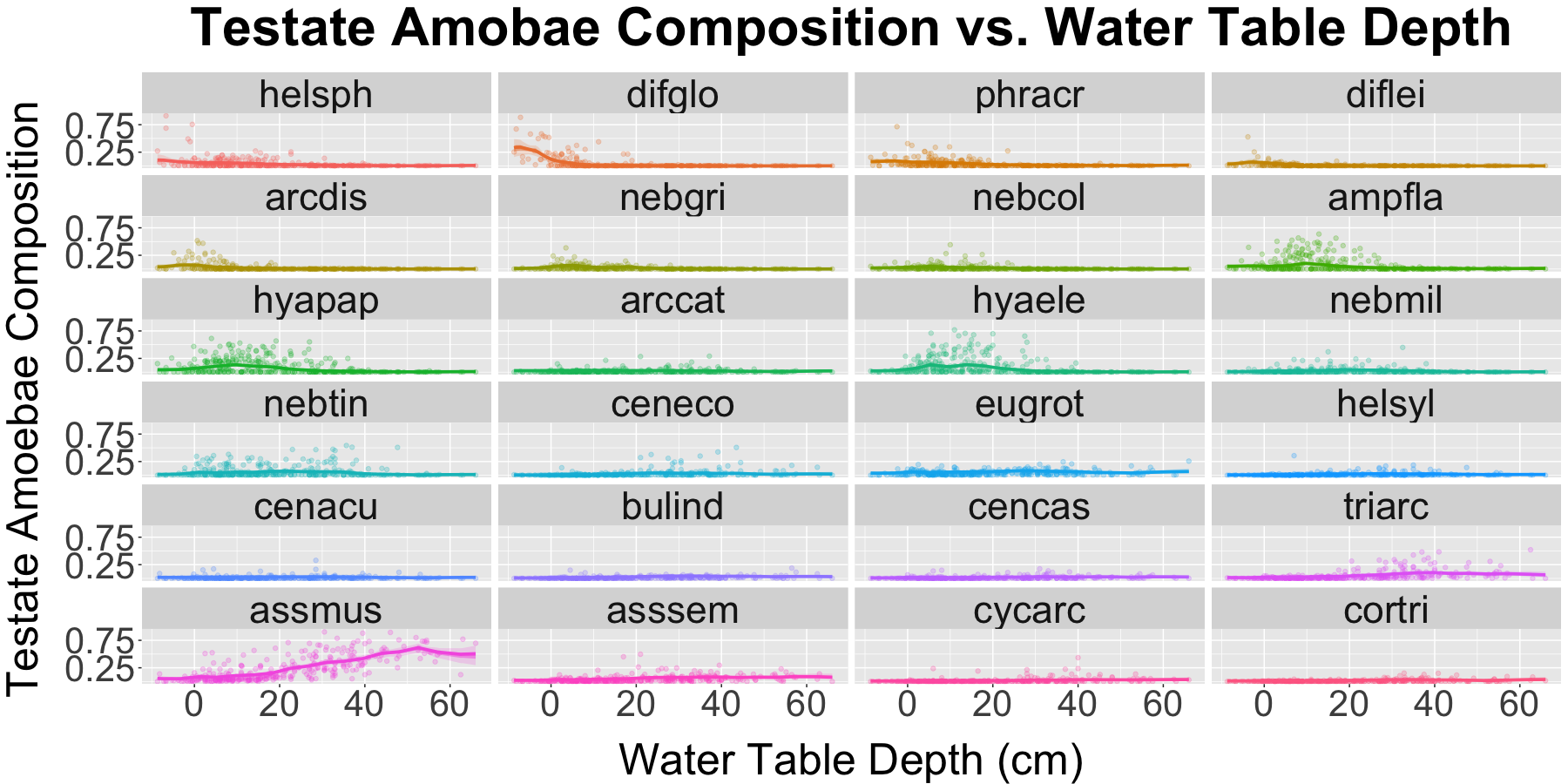}
  \caption{Posterior mean response.}
  \label{fig:testate-fit}
\end{subfigure}%
\begin{subfigure}{0.5\textwidth}
  \centering
  \includegraphics[width=1.0\linewidth]{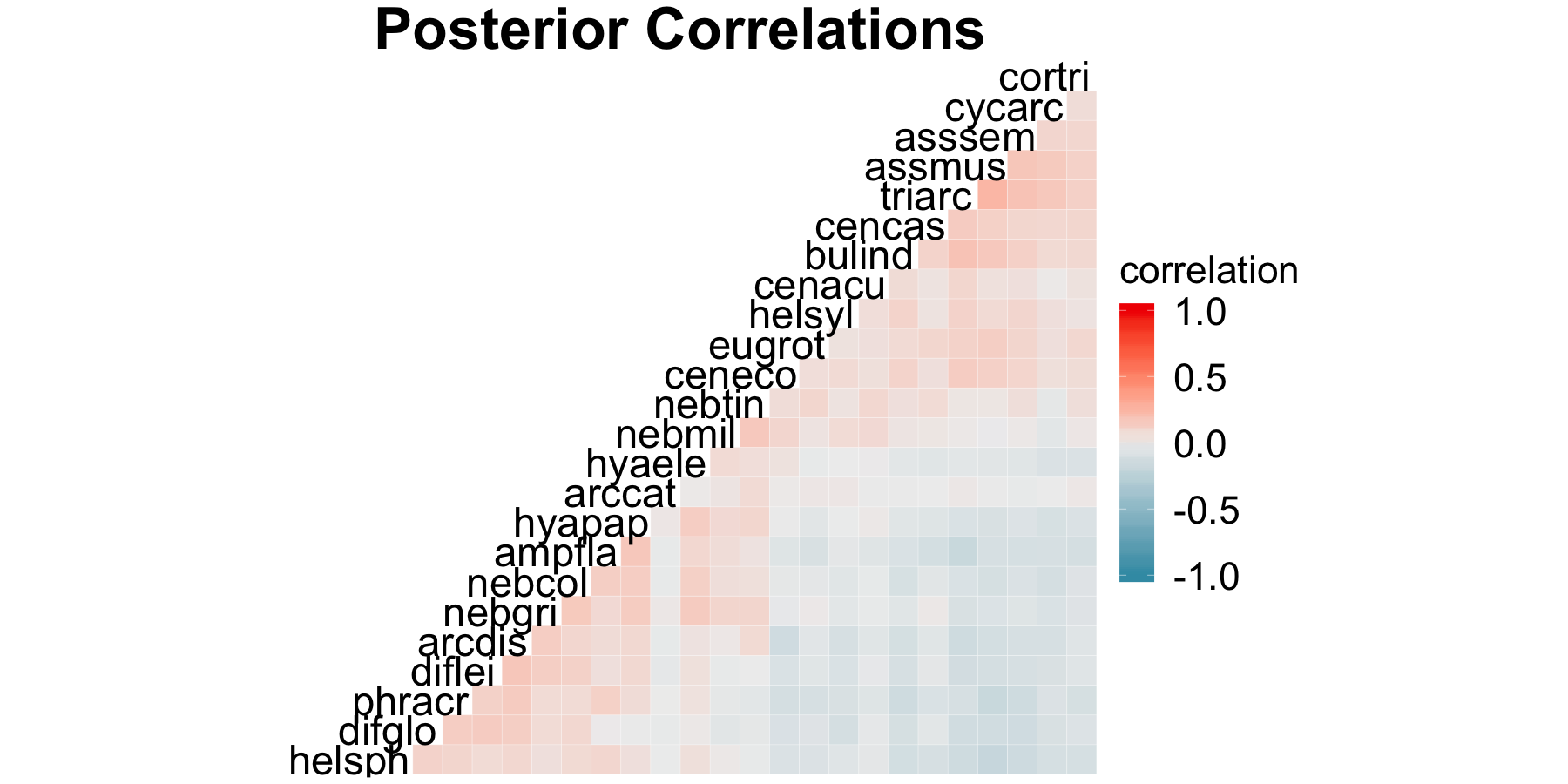}
  \caption{Posterior Correlations.}
  \label{fig:testate-corr}
\end{subfigure}
\caption{Posterior model fit for the testate amoeba data. Figure (a) shows the fitted latent response of each species' contribution to the composition with respect to water table depth, sorted by increasing maximal species response to water table depth from top left to bottom right. Figure (b) shows the posterior mean pairwise correlations in response to water table depth among testate amoeba species, sorted by increasing maximal species response to water table depth from bottom left to top right. Red colors show positive correlations and blue colors show negative correlation.}
\end{figure}

\begin{figure}
\centering
\begin{subfigure}{0.5\textwidth}
  \centering
  \includegraphics[width=1.0\linewidth]{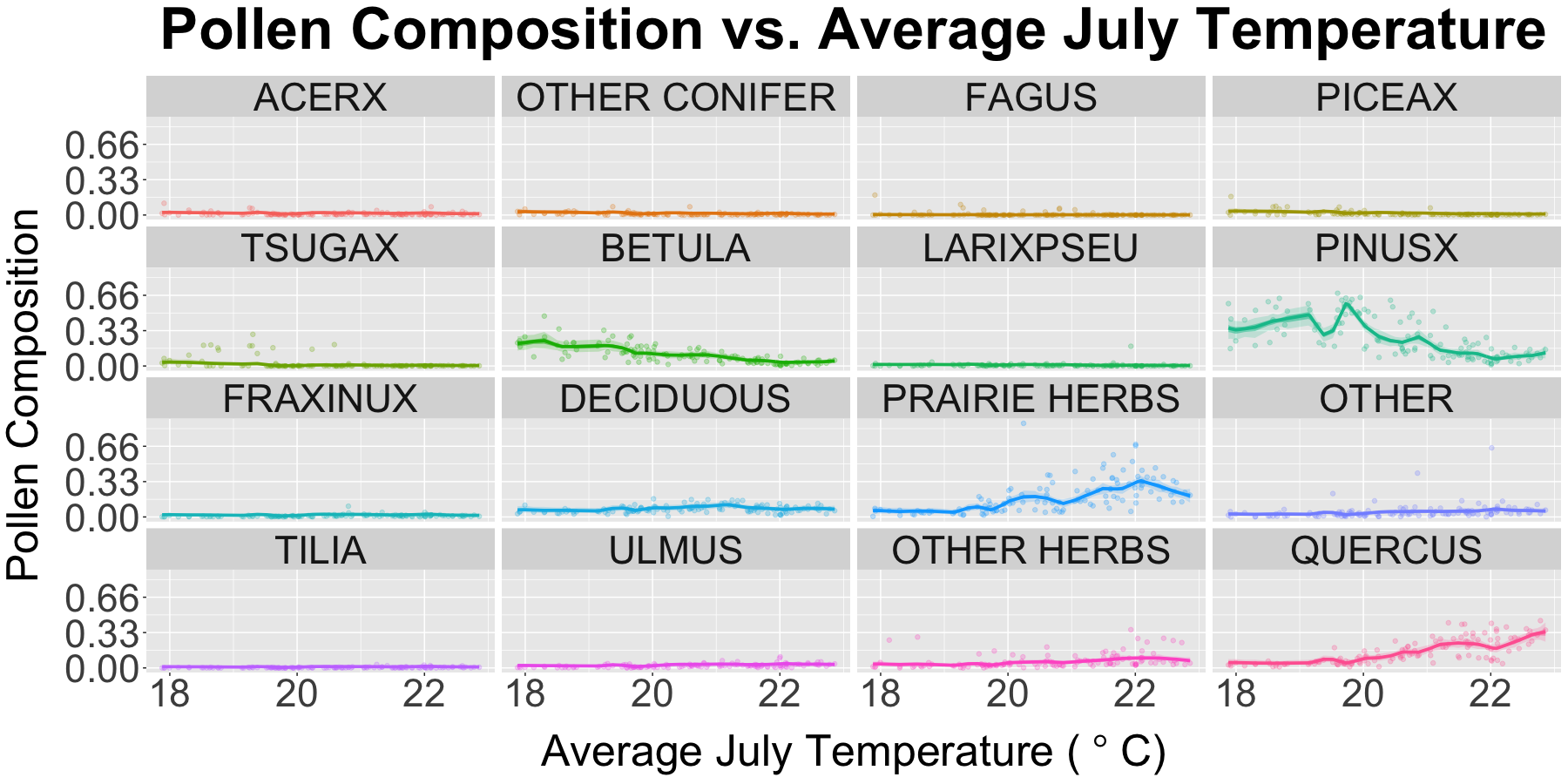}
  \caption{Posterior mean response.}
  \label{fig:pollen-fit}
\end{subfigure}%
\begin{subfigure}{0.5\textwidth}
  \centering
  \includegraphics[width=1.0\linewidth]{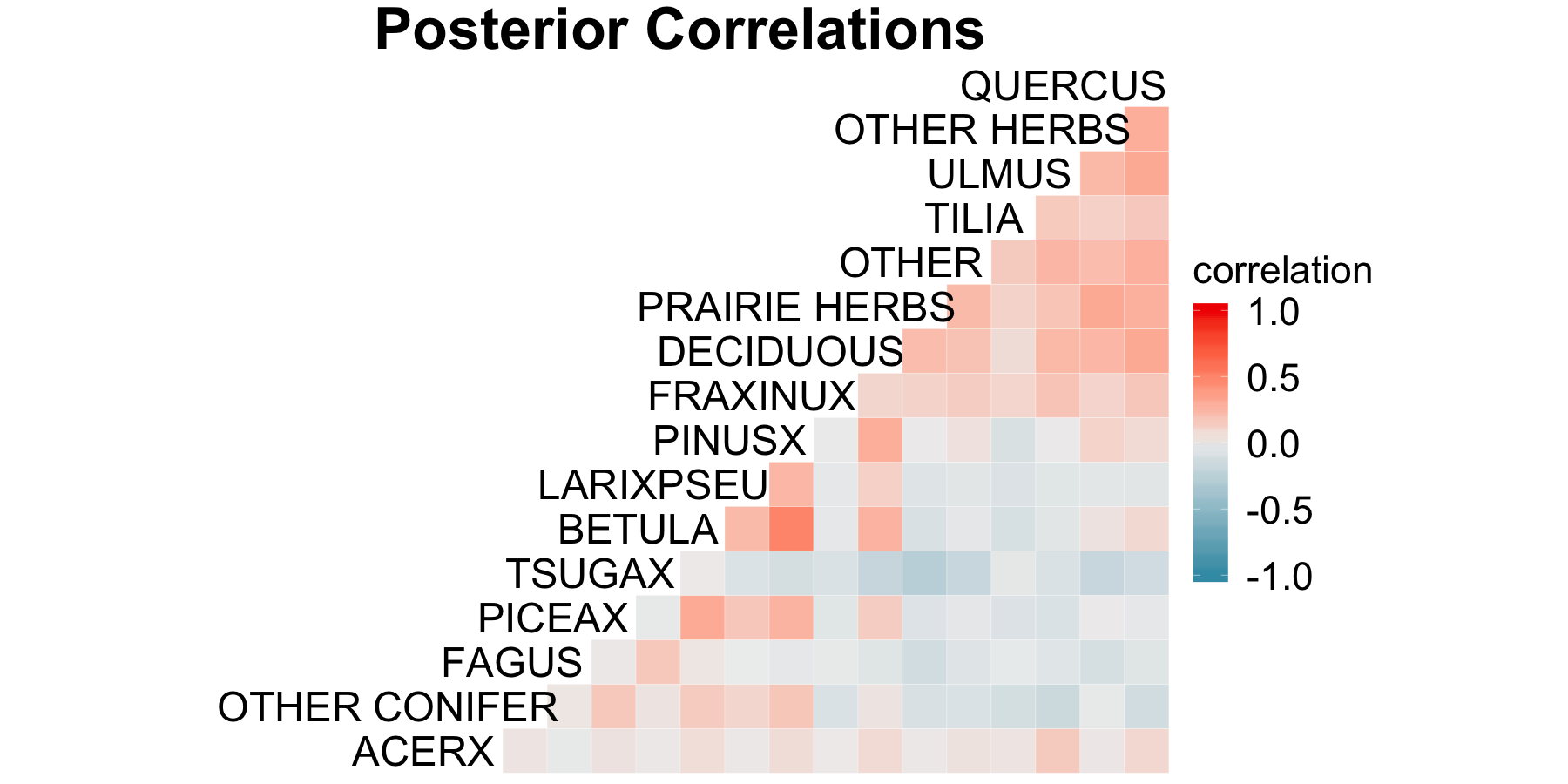}
  \caption{Posterior Correlations.}
  \label{fig:pollen-corr}
\end{subfigure}
\caption{Posterior model fit for the pollen data. Figure (a) shows the fitted latent response of each species' contribution to the composition with respect to average July temperature, sorted by increasing maximal species response to temperature from top left to bottom right. Figure (b) shows the posterior mean pairwise correlations in response to average July temperature among species, sorted by increasing maximal species response to average July temperature from bottom left to top right. Red colors show positive correlations and blue colors show negative correlation.}
\end{figure}

The MVGP, BUMMER, and GAM models also allow for inference that is not
available with the transfer function methods (WA, MAT, and MLRC). The
ability to make inference on the latent functional relationship between
composition data and unobserved covariates is valuable. Figures
\ref{fig:testate-fit} and \ref{fig:pollen-fit} show the MVGP posterior
mean response of each testate and pollen species to water table depth
and average July temperature, respectively. Using these figures,
researchers can make meaningful inference about the ecological niche
that different species are exploiting. For example, in Figure
\ref{fig:testate-fit} the species \emph{Assulina muscorum} (assmus) is
dominant in environments with water table depths deep below the peatland
surface while the species \emph{Hyalosphenia elegans} (hyaele) is more
prevalent when water tables are near the surface. In Figure
\ref{fig:pollen-fit}, there is a pronounced multimodal response of
\emph{Pinus} species to average July temperature. The \emph{Pinus} taxa
is a combination of different pine species that each have a different
ecological niche and would be expected to have a multimodal functional
response that would not be detected under the BUMMER model.

The posterior mean estimates of the correlations among functional
responses shown in Figure \ref{fig:testate-corr} and Figure
\ref{fig:pollen-corr} provide insight into potential interactions among
species. For example, some members of the same genus, such as
\emph{Nebula} (neb) and \emph{Assulina} (ass), show positive
correlations in their functional response to water depth. Not
surprisingly though, there are a number of unrelated species that show
high correlations, such as \emph{Trigonopyxis arcula} (triarc) and
\emph{Assulina spp} (assmus, asssem), likely because they occupy similar
niches with respect to surface moisture and other environmental
conditions. This is supported in Figure \ref{fig:testate-corr} by the
presence of positive correlations near the diagonal of the correlation
matrix. The positive correlations near the diagonal in Figure
\ref{fig:testate-corr} suggest species that have maximal responses that
are near each other display a correlation in the functional response to
water table depth. Importantly, species with similar energetic
strategies, in particular mixotrophic species like \emph{Hyalosphenia
papilio} (hyapap) and \emph{Heleopera sphagni} (helsph) that contain
endosymbiotic zoochorellae also have high correlations, likely due to
similar water-table depth tolerances as well as light requirements.
Likewise Figure \ref{fig:pollen-corr} shows the positive correlations of
plant species with maximal responses at either the cool or hot end of
the average July temperature gradient. For example, \emph{Pinus} and
\emph{Betula} are highly correlated in their functional response, which
is not surprising as these taxa frequently occur together. Species with
high correlations could be clustered together in future statistical
analysis to reduce the number of parameters that need to be estimated
and might also save analysts time by reducing the need for taxonomic
precision in some cases. These inferential questions cannot be answered
by the transfer function techniques commonly used to model compositional
data, demonstrating the value of the MVGP modeling approach.

\section{Discussion}

This work developed a flexible, novel model framework for prediction of
unobserved climate from compositional count data. We have shown the MVGP
model is capable of providing predictions that are of equal or superior
skill to current methods depending on the properties of the underlying
data. Although MVGP might not always be the most skilled model in
cross-validation using the observed data, the MVGP model is robust to
the issues commonly seen in using compositional count data to
reconstruct climate including the ``no-analog'' setting. In
``no-analog'' data, the cross-validation scores for MVGP dominate the
other models (Appendix S5). Thus, although MVGP does not uniformly
outperform WA and MAT in cross-validation, the empirical robustness of
MVGP to the ``no-analog'' problem (Appendix S5) and performance on data
simulated from the BUMMER model suggests that our model framework is
more robust to data seen in practice.

We have shown that the MVGP model can be a useful inferential tool by
fitting the model to simulated data and showing that the inference
accurately reproduces simulated parameters. This is important because
posterior inference that we obtain from the MVGP provides learning about
the underlying processes that is not available in the transfer function
methods WA and MAT while making fewer assumptions than the current
Bayesian methodology of BUMMER. The learning about the correlations in
functional responses can be used to validate the groupings of taxa into
similar functional types and provide the basis for a data-driven method
of grouping taxa based on similar functional responses. Using the
learning of the underlying processes giving rise to the data, we can
guide future model and dataset development to iteratively improve
paleoclimate reconstructions.

The statistical and computational framework developed in this manuscript
allowed us to fit a complex model to compositional count data in a
reasonable amount of time. We presented an algorithm for predicting
unobserved inputs into a Gaussian process model by resolving a
computational bottleneck that allowed for use of the highly flexible
Gaussian process functional form to be used for inverse prediction. We
also implemented an MCMC sampling technique that can efficiently explore
the multimodal posterior generated by the inverse prediction framework.

Finally, MVGP and the other probabilistic methods for climate
reconstruction have greater flexibility in future development. It is
natural to extend the probabilistic models to include temporal or
spatial autocorrelation by assigning the unobserved covariates
\(\mathbf{X}\) a correlated prior. Possible autocorrelation structures
can account for either continuous or discrete observations in space and
time. In addition, because the likelihood is probabilistic, one can
account for radio-dating uncertainty through weighting the likelihoods
using an estimated age-depth model to more fully account for
uncertainty. The impact of these changes is difficult to include in a
cross-validation experiment but better incorporates domain knowledge of
how the climate processes evolve in time and space. Given that the MVGP
model is competitive in predictive skill with current modeling efforts,
the ability to extend MVGP to correlated spatio-temporal reconstructions
is a great benefit. Additional improvements in reconstruction skill
could also be seen by combining MVGP with other proxy data sources to
provide more precise inference of paleoclimate.

\section{Acknowledgements}

This research is based upon work carried out by the PalEON Project
(paleonproject.org) with support from the National Science Foundation
MacroSystems Biology program under grant no. DEB-1241856. Any use of
trade, firm, or product names is for descriptive purposes only and does
not imply endorsement by the U.S. Government. Code and data found in
this manuscript can be accessed at
\url{http://github.com/jtipton25/compositional-inverse-prediction}.

\bibliographystyle{imsart-nameyear}\bibliography{multivariate-gaussian-process.bib}

\end{document}